\documentclass[10pt,twocolumn]{article}

\usepackage[utf8]{inputenc}
\usepackage{amsmath}
\usepackage{amsthm}
\usepackage{amssymb}
\usepackage{titlesec}
\usepackage{cite}
\usepackage{booktabs}
\usepackage{graphicx}
\usepackage[colorlinks=false]{hyperref}
\usepackage[format=plain,labelfont=it]{caption}
\usepackage[left=1.5cm,right=1.5cm,top=2cm,bottom=2cm]{geometry}
\usepackage{color}

\titleformat{\section}{\centering\normalfont\scshape}{\Roman{section}.}{5pt}{}
\titleformat{\subsection}{\normalfont\it}{\Alph{subsection}.}{5pt}{}
\titleformat{\subsubsection}{\normalfont\it}{\hspace{4mm}\arabic{subsubsection})}{5pt}{}

\newcommand\infoFootnote[1]{%
  \begingroup
  \renewcommand\thefootnote{}\footnote{#1}%
  \addtocounter{footnote}{-1}%
  \endgroup}

\usepackage{tikz}
\usetikzlibrary{shapes,arrows, arrows.meta}
\usetikzlibrary{positioning}

\tikzstyle{block} = [draw, fill=gray!20, rectangle, 
    minimum height=3em, minimum width=6em]

\tikzset{
pics/mynode/.style args={}{
 code={
   \draw[blue] (0,0) coordinate (Origin) --++(0.2,0) coordinate (endL) --++(0.1,0) --++(0.2,0) coordinate (endR) --++(45:0.25) arc (180-45:0:0.08) (endR) arc (0:-180:0.1) (endL) arc (0:-180:0.1) (Origin)--++(45:0.25) arc (180-45:0:0.08);
 }
}
}

\DeclareFontFamily{U}{mathx}{}
\DeclareFontShape{U}{mathx}{m}{n}{<-> mathx10}{}
\DeclareSymbolFont{mathx}{U}{mathx}{m}{n}
\DeclareMathAccent{\widehat}{0}{mathx}{"70}
\DeclareMathAccent{\widecheck}{0}{mathx}{"71}

\newtheorem{theorem}{Theorem}
\newtheorem{corollary}[theorem]{Corollary}

\newtheorem{proposition}[theorem]{Proposition}
\newtheorem{assumption}{Assumption}

\newtheorem{definition}{Definition}

\newtheorem{remark}{Remark}

\makeatletter
\renewcommand*\env@matrix[1][\arraystretch]{%
  \edef\arraystretch{#1}%
  \hskip -\arraycolsep
  \let\@ifnextchar\new@ifnextchar
  \array{*\c@MaxMatrixCols c}}
\makeatother

\DeclareMathOperator*{\argmin}{\arg\min}

\newcommand{\R}{\mathbb{R}} 

\newcommand{\ab}{\boldsymbol{a}}

\newcommand{\eb}{\boldsymbol{e}}

\newcommand{\gb}{\boldsymbol{g}}
\newcommand{\kb}{\boldsymbol{k}}
\newcommand{\rb}{\boldsymbol{r}}
\newcommand{\ub}{\boldsymbol{u}}
\newcommand{\wb}{\boldsymbol{w}}
\newcommand{\xb}{\boldsymbol{x}}
\newcommand{\yb}{\boldsymbol{y}}
\newcommand{\xib}{\boldsymbol{\xi}}
\newcommand{\sigmab}{\boldsymbol{\sigma}}
\newcommand{\zerob}{\boldsymbol{0}}
\newcommand{\oneb}{\boldsymbol{1}}

\newcommand{\Ab}{\boldsymbol{A}}
\newcommand{\Bb}{\boldsymbol{B}}
\newcommand{\Cb}{\boldsymbol{C}}
\newcommand{\Db}{\boldsymbol{D}}
\newcommand{\Eb}{\boldsymbol{E}}
\newcommand{\Gb}{\boldsymbol{G}}

\newcommand{\Ib}{\boldsymbol{I}}
\newcommand{\Kb}{\boldsymbol{K}}
\newcommand{\Lb}{\boldsymbol{L}}

\newcommand{\Qb}{\boldsymbol{Q}}

\newcommand{\Sb}{\boldsymbol{S}}
\newcommand{\Ub}{\boldsymbol{U}}

\newcommand{\Wb}{\boldsymbol{W}}
\newcommand{\Xb}{\boldsymbol{X}}
\newcommand{\Yb}{\boldsymbol{Y}}
\newcommand{\Zb}{\boldsymbol{Z}}
\newcommand{\zb}{\boldsymbol{z}}
\newcommand{\Pib}{\boldsymbol{\Pi}}

\newcommand{\dYb}{\Eb_\ybs}
\newcommand{\dUb}{\Eb_\ubs}
\newcommand{\dyb}{\eb_\ybs}
\newcommand{\dub}{\eb_\ubs}

\newcommand{\gammab}{\boldsymbol{\gamma}}

\newcommand{\ybs}{\mathbf{y}}
\newcommand{\xbs}{\mathbf{x}}
\newcommand{\ubs}{\mathbf{u}}

\newcommand{\Dbc}{\boldsymbol{\mathcal{D}}}
\newcommand{\Obc}{\boldsymbol{\mathcal{O}}}
\newcommand{\Qbc}{\boldsymbol{\mathcal{Q}}}
\newcommand{\Rbc}{\boldsymbol{\mathcal{R}}}

\newcommand{\Nc}{\mathcal{N}}
\newcommand{\Uc}{\mathcal{U}}
\newcommand{\Xc}{\mathcal{X}}
\newcommand{\Yc}{\mathcal{Y}}

\newcommand{\image}[1]{\mathcal{R}\left({#1}\right)}
\newcommand{\rank}[1]{\mathrm{rank}\left({#1}\right)}

\newcommand{\diag}[1]{\mathrm{diag}\left({#1}\right)}

\renewcommand{\boldsymbol}[1]{#1}
\renewcommand{\mathbf}[1]{\mathrm{#1}}

\title{\vspace{-2mm}\bf Towards explainable data-driven \\ predictive control with regularizations}
\author{Manuel Kl\"adtke and Moritz Schulze Darup \vspace{2mm}}
\date{}

\begin{document}
\maketitle

\begin{abstract}
Data-driven predictive control (DPC), using linear combinations of recorded trajectory data, has recently emerged as a popular alternative to traditional model predictive control (MPC). Without an explicitly enforced prediction model, the effects of commonly used regularization terms—and the resulting predictions—can be opaque. This opacity may lead to practical challenges, such as reliance on empirical tuning of regularization parameters based on closed-loop performance, and potentially misleading heuristic interpretations of norm-based regularizations. However, by examining the structure of the underlying optimal control problem (OCP), more precise and insightful interpretations of regularization effects can be derived.
In this paper, we demonstrate how to analyze the predictive behavior of DPC through implicit predictors and the trajectory-specific effects of quadratic regularization. We further extend these results to cover typical DPC modifications, including DPC for affine systems, offset regularizations, slack variables, and terminal constraints. Additionally, we provide a simple but general result on (recursive) feasibility in DPC. This work aims to enhance the explainability and reliability of DPC by providing a deeper understanding of these regularization mechanisms.
\end{abstract}
% leave no space here
\infoFootnote{M. Kl\"adtke and M. Schulze Darup are with the \href{https://rcs.mb.tu-dortmund.de/}{Control and~Cyberphysical Systems Group}, Faculty of Mechanical Engineering, TU Dortmund University, Germany. E-mails:  \href{mailto:manuel.klaedtke@tu-dortmund.de}{\{manuel.klaedtke, moritz.schulzedarup\}@tu-dortmund.de}. \vspace{0.5mm}}
\infoFootnote{This is an Original Manuscript of an article to be published by De Gruyter in \textit{at - Automatisierungstechnik}.} 

\section{Introduction} 
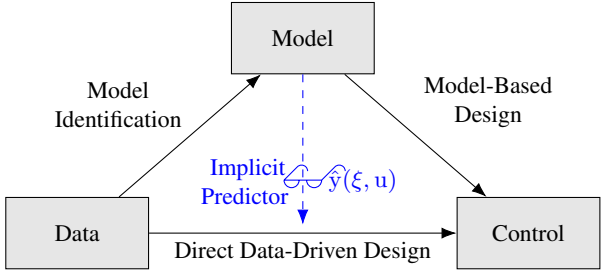
\begin{figure}
    \centering
    {\small% \documentclass{article}

% \usepackage{tikz}
% \usetikzlibrary{shapes,arrows}
% \usetikzlibrary{positioning}

% \begin{document}

% \tikzstyle{block} = [draw, fill=gray!20, rectangle, 
%     minimum height=3em, minimum width=6em]

% \tikzset{
% pics/mynode/.style args={}{
%  code={
%    \draw (0,0) coordinate (Origin) --++(0.2,0) coordinate (endL) --++(0.1,0) --++(0.2,0) coordinate (endR) --++(45:0.25) arc (180-45:0:0.08) (endR) arc (0:-180:0.1) (endL) arc (0:-180:0.1) (Origin)--++(45:0.25) arc (180-45:0:0.08);
%  }
% }
% }

\begin{tikzpicture}[auto, node distance=3cm,>=latex']
    \node [block] (data) {Data};
    \node [right of= data] (placeholder) {};
    \node [block, right of= placeholder] (control) {Control};
    \node [block, above =2cm of placeholder] (model) {Model};
    \node [above =0cm and 0.5cm of placeholder] (predictor) {};
    \draw [-{Latex[length=2mm]}] (data) -- node[name=sysID, align=center] { Model \\ Identification } (model);
    \draw [-{Latex[length=2mm]}] (model) -- node[name=indirect, align=center] { Model-Based \\ Design } (control);
    \draw [-{Latex[length=2mm]}] (data) -- node[below, name=direct] {Direct Data-Driven Design} (control);
    \draw [dashed, -{Latex[length=2mm]}, blue] (model) -- node[left, near end, name=predictorLabel, align=center, yshift=0.08cm] {\color{blue} Implicit \\ \color{blue}Predictor }  (placeholder);
    \node [label={[xshift=0.8cm, yshift=0.08cm]\color{blue} $\hat\ybs(\xib, \ubs)$}] at (predictor) {  };
    \draw (predictor) +(-0.25, 0.48) pic{ mynode={}} {};
    
\end{tikzpicture}

%\end{document}
}\vspace{-1cm}
    \caption{Direct data-driven control schemes aim to design control directly from data. This is in contrast to the indirect (i.e., model-based) data-driven control design paradigm. Here,  we aim for an indirect viewpoint (highlighted in blue) on the predictions made by direct schemes via implicit predictors.
    }\label{fig:implicitPredictor}
\end{figure}

Data-driven predictive control (DPC,  \cite{Yang2013, Coulson2019DeePC, Berberich2020stability}), utilizing linear combinations of recorded trajectory data to make predictions, rather than relying on an explicit system model, has recently emerged as a popular alternative to classical model predictive control (MPC, \cite{Rawlings2017}). This control paradigm exemplifies a direct data-driven control scheme, contrasting with indirect (model-based) methods, as visualized in Figure~\ref{fig:implicitPredictor}. Originally, DPC is theoretically grounded in a result from behavioral system theory \cite{WILLEMS2005}, known as the Willems' Fundamental Lemma. Furthermore, DPC yields an exact system representation and equivalence to MPC if the underlying system is deterministic LTI \cite{Coulson2019DeePC, Berberich2020}. While this exactness extends to certain classes of deterministic nonlinear systems \cite{Berberich2020, Alsalti2021}, it is typically lost in realistic cases involving noise, disturbances, and general nonlinearities. To address this, regularization terms are often added to the DPC objective function \cite{Coulson2019DeePC}, initially motivated by their relation to distributional robustness \cite{Coulson2019RegularizedDeePC}. Since then, regularized DPC schemes have shown promising performance, and further theoretical justification via robustness and stability results \cite{Berberich2020stability, Berberich2021, Berberich2022Tutorial, alsalti2024robustDPC}. However, the heuristic nature of regularizations often obscures their direct impact on the synthesis of predicted system trajectories from data, and since DPC operates without an explicitly enforced prediction model, the precise influence of regularizations on the resulting predictions can be challenging to discern. Therefore, our aim is to provide a deeper analysis that clarifies the interaction of control objective, constraints, and regularizations in DPC. Specifically, we propose two tools for this analysis, namely the trajectory-specific effect of regularizations (see Definition~\ref{def:trajectorySpecific}) and implicit predictors (see Definition~\ref{def:implicit_predictor}), first introduced in \cite{KLAEDTKE2023}. The former reformulates regularization costs, translating their effect from auxiliary variables to the actual predicted system variables. The latter yields a model-based perspective on DPC by generating prediction mappings that align with DPC's actual predictions (see Figure~\ref{fig:implicitPredictor}). We demonstrate the use of these analysis tools by summarizing previous results from \cite{KLAEDTKE2023, Klaedtke2024FCS_DPC, klaedtke2024unifyingPreprint} and extend them towards common modifications in DPC. The paper is organized as follows. First, in Section~\ref{sec:PreliminariesExample} we summarize fundamentals on direct data-driven predictions and regularized DPC, and discuss a numerical example, which is used for visualization throughout the paper. In Section~\ref{sec:motivation}, we further motivate our approach and explain its underlying assumptions. Section~\ref{sec:trajectory-specific} presents the trajectory-specific effect of regularization and summarizes related findings from \cite{KLAEDTKE2023, Klaedtke2024FCS_DPC, klaedtke2024unifyingPreprint}. In Section~\ref{sec:implicitPredictors}, we introduce implicit predictors and demonstrate their use by summarizing results from \cite{KLAEDTKE2023, Klaedtke2024FCS_DPC, klaedtke2024unifyingPreprint}. Section~\ref{sec:extensions} expands these results to common DPC modifications, including affine systems, offset regularization, slack variables, and (terminal) equality constraints, and provides a general feasibility result. Finally, we conclude our work in Section~\ref{sec:ConclusionOutlook} and preview future applications for our proposed analysis tools. 

\section{Preliminaries and running example}\label{sec:PreliminariesExample}

\subsection{Fundamentals of direct data-driven predictions} \label{sec:fundamentals}
\begin{figure}
    \centering
    \includegraphics[trim=5.6cm 12.4cm 6cm 12.4cm,clip=true, scale=0.88]{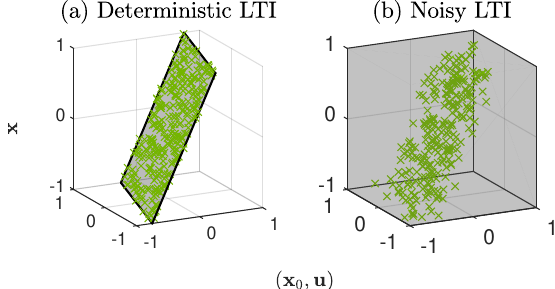}
        \caption{Visualization of the data matrix $\Dbc$ for the low-dimensional state-space example in Section~\ref{sec:example}. The green marks show data columns $\left(\xb_0^{(i)}, \ubs^{(i)}, \xbs^{(i)}\right)$ for (a) the ideal deterministic LTI setting and (b) the same data with added measurement noise. The gray subspaces depict $\image{\Dbc}$, showing $\rank{\Dbc}=\rank{\Zb}=2$ in (a) and full rank $\Dbc$ with $\rank{\Dbc}=3>\rank{\Zb}=2$ in (b).}
        \label{fig:visualizeD}
\end{figure}

Instead of utilizing a discrete-time state-space model with input $\ub\in\R^m$, state $\xb\in\R^n$, and output ${\yb\in\R^p}$ as in traditional MPC, predictions in DPC are realized based on previously collected trajectory data $\big(\ubs^{(1)}, \ybs^{(1)}\big), \hdots, \big(\ubs^{(\ell)}, \ybs^{(\ell)}\big)\in\R^{m L}\times \R^{p L}$ via linear combinations
\begin{equation}
    \begin{pmatrix}
        \ubs_\text{gen}\\
        \ybs_\text{gen}
    \end{pmatrix}
    =
    \begin{pmatrix}
        \ubs^{(1)}\\
        \ybs^{(1)}
    \end{pmatrix} a_1 + \hdots + 
    \begin{pmatrix}
        \ubs^{(\ell)}\\
        \ybs^{(\ell)}
    \end{pmatrix} a_\ell
    = 
    \Dbc
    \ab. \label{eq:linearComb}
\end{equation}
Here, the dimensions of the data matrix $\Dbc\in\R^{L(m+p)\times\ell}$ and generator vector $\ab\in\R^\ell$ are specified by the length~$L$ of recorded (and generated) trajectories and the number~$\ell$ of data trajectories used for predictions. Furthermore, we refer to a pair $\big(\ubs, \ybs\big)$ as an input/output-sequence (\textit{I/O-sequence}), and use roman notation to distinguish individual variables like $\ub, \yb$ from their associated sequences $\ubs, \ybs$. Assuming exact data generated by an LTI system, which can be described, e.g., by the state-space representation
\begin{subequations}\label{eq:statespace_LTI}
\begin{align}
    \xb (k+1) &= \Ab\xb(k)+\Bb\ub(k)\\
    \yb(k) &= \Cb\xb(k)+\Db\ub(k),
\end{align}
\end{subequations}
and assuming $L$ is greater than the lag $\iota$ of the system, the image $\image{\Dbc}$ is equivalent to the set of all possible system trajectories (of length $L$) if and only if \cite{Markovsky2020}
\begin{equation}\label{eq:GPE} 
    \rank{\Dbc}=L m + n. 
\end{equation} 
Note that the lag $\iota$ is an integer invariant \cite{WILLEMS2005} of the system, i.e., invariant with respect to the considered representation. In context of the state-space representation \eqref{eq:statespace_LTI}, it is also known as its observability index, which is defined as the smallest integer $\iota$, for which $\rank{\Obc_\iota}=\rank{\Obc_{\iota+1}}$ with
$$
    \Obc_\iota := \begin{pmatrix}
        \Cb \\
        \Cb \Ab \\
        \vdots \\
        \Cb \Ab^{\iota-1}
    \end{pmatrix}
$$
is satisfied.
The generalized persistency of excitation condition \eqref{eq:GPE} provides the theoretical foundation for the linear combinations \eqref{eq:linearComb}, since 
\begin{equation}
    \begin{pmatrix}
        \ubs_\text{gen}\\
        \ybs_\text{gen}
    \end{pmatrix}
    \in
    \image{\Dbc} \; \iff \; \exists \ab\; \text{such that}\; 
    \begin{pmatrix}
        \ubs_\text{gen}\\
        \ybs_\text{gen}
    \end{pmatrix} = \Dbc\ab. \label{eq:imageRep0}
\end{equation}
Representing system trajectories in this way is also known as an image representation of the system, as opposed to, e.g., a state-space representation \eqref{eq:statespace_LTI}. A popular sufficient condition for data to satisfy \eqref{eq:GPE} is known as Willems' Fundamental Lemma \cite{WILLEMS2005}, which has become synonymous with using image representations. We note that, contrary to the Fundamental Lemma, the generalized persistency of excitation condition \eqref{eq:GPE} is both sufficient and necessary, and neither requires controllability nor a Hankel structure for the data matrix.
To include the current initial condition of the system as a starting point for predicted trajectories, the generated I/O-sequence is typically partitioned into a past section $(\ubs_p, \ybs_p)$ and a future section $(\ubs_f, \ybs_f)$ with $N_p$ respectively $N_f$ time-steps yielding
$$
    \begin{pmatrix}
        \ubs_p\\
        \ubs_f
    \end{pmatrix} = \ubs_\text{gen} 
    = 
    \begin{pmatrix}
        \Ub_p\\
        \Ub_f
    \end{pmatrix} \ab \quad \text{and} \quad
    \begin{pmatrix}
        \ybs_p\\
        \ybs_f
    \end{pmatrix} = \ybs_\text{gen} 
    = 
    \begin{pmatrix}
        \Yb_p\\
        \Yb_f
    \end{pmatrix} \ab. 
$$
The past section of a predicted trajectory is then forced to match the I/O-data $\xib$ recorded in the most recent $N_p$ time-steps during closed-loop operation, i.e., the constraints
$$
    \xib= 
    \begin{pmatrix}
        \ubs_p \\ \ybs_p
    \end{pmatrix}
    =
    \begin{pmatrix}
        \Ub_p \\ \Yb_p
    \end{pmatrix}\ab
    = \Wb \ab
$$
force any predicted trajectory to start with the most recently witnessed behavior of the system. Note that $\xib$ is a (non-minimal) state of the LTI system \eqref{eq:statespace_LTI}, if $N_p$ is chosen greater or equal to its lag $\iota$. From now on, we omit the ``future'' subscript from $\ubs_f, \ybs_f, \Ub_f$, and $\Yb_f$, since their ``past'' counterparts are already incorporated in $\Wb$ and $\xib$, eliminating any risk of confusion. For more concise notation, we also define 
$$\Zb := \begin{pmatrix}
        \Wb^\top & \Ub^\top
    \end{pmatrix}\!^\top,\,\,  \zb = \begin{pmatrix}
        \xib^\top & \ubs^\top
    \end{pmatrix}\!^\top,\,\,   \text{and}\,\,  \wb = \begin{pmatrix}
        \zb^\top & \ybs^\top
    \end{pmatrix}\!^\top\!.$$
    Additionally, with a slight abuse of notation, we redefine
    $\Dbc := \big(
        \Zb^\top \quad \Yb^\top
\big)^\top$.
The latter is simply a block-row permuted version of the original data matrix introduced in~\eqref{eq:linearComb}, used to facilitate the corresponding partitioning of $\ubs_\text{gen}, \ybs_\text{gen}$ into $\xib, \ubs, \ybs$. To summarize these new notations, $\Wb$ contains data columns for the state $\xib$, $\Ub$ contains data columns for the future input sequence $\ubs$, and $\Yb$ contains data columns for the future output sequence $\ybs$. Furthermore, $\Zb$ contains data for the state-input sequence $\zb$, and $\Dbc$ contains data for the state-input-output sequence $\wb$. 
\begin{remark}
    \label{rem:Statespace}
    Although we have introduced the data-driven predictions in an I/O setting, they can be straightforwardly modified to a state-space setting \cite{DePersis2020}.  To this end, consider
    \begin{align*}
        \xib\, &\hat = \, \xb_0\in\R^n, &\Wb\, &\hat =\, \Xb_0\in\R^{n\times\ell}, \\ \ybs \, &\hat=\, \xbs\in \R^{n N_f},\qquad \text{and} \;&\Yb \, &\hat = \, \Xb\in\R^{n N_f\times \ell}
    \end{align*}
    leading to a data matrix $\Dbc$ of dimension $n+(m+n)N_f\times \ell$.
    Our considerations and condition \eqref{eq:GPE} with $\rank{\Dbc}=n+m N_f$ hold for both settings, and we employ     the latter for visualization of a low dimensional numerical example used throughout the paper and 
   detailed in Section~\ref{sec:example}.
\end{remark}
Notably, in the ideal deterministic LTI setting with condition \eqref{eq:GPE}, the data matrix always has the rank deficiency
\begin{equation}
    \rank{\Dbc} = \rank{\Zb}, \label{eq:rankDef}
\end{equation}
such that the image representation \eqref{eq:imageRep0} implies a unique (and exact) linear predictor mapping $\ybs = \hat\ybs(\xib, \ubs)$.
However, this rank deficiency, and with it the unique predictions, are typically lost in the presence of noise or nonlinearities, which is visualized in Figure~\ref{fig:visualizeD}.
One remedy for this might be the use of (structured) low-rank approximations \cite{MARKOVSKY2008surveySLRA} for $\Dbc$, which sometimes even implicitly occurs in DPC schemes, e.g., in $\gammab$-DDPC \cite{Breschi2022new} with $\gammab_3 = \zerob$ (see \cite{klaedtke2024unifyingPreprint}). Specific to the context of DPC, another remedy is given by the addition of a regularization term $h(\ab)$ to the objective function \eqref{eq:DPCcost} \cite{Coulson2019DeePC, Berberich2020stability}, which
is in
the focus of this work.

\subsection{Regularized DPC}

The optimal control problem (OCP) that is solved for DPC in every time step can be stated as 
\begin{subequations}
\label{eq:DPC}
\begin{align}
\min_{\ubs,\ybs,\ab} 
 J(\xib, \ubs, &\ybs) + h(\ab) \label{eq:DPCcost} \\
\text{s.t.} \quad \quad  \begin{pmatrix}
     \xib \\ \ubs \\ \ybs 
\end{pmatrix} &= \begin{pmatrix}
    \Wb \\ \Ub \\ \Yb 
\end{pmatrix}\ab, \label{eq:DPCeqConstr}\\
\left(\ubs, \ybs \right) &\in \Uc \times \Yc.  \label{eq:DPCsetConstr}
\end{align}
\end{subequations}
with control objective $J(\xib, \ubs, \ybs)$, regularization $h(\ab)$, and input-output constraints $\Uc \times \Yc$. Conditions for equivalence of DPC and MPC are well established for some special cases. In particular, as mentioned in the introduction, the equivalence with MPC for the LTI system \eqref{eq:statespace_LTI} holds with exact data satisfying \eqref{eq:GPE} and setting the regularization to $h(\ab) = 0$ \cite{Coulson2019DeePC}. In the presence of noise or nonlinearities, the unregularized OCP may use the (unrealistic) additional degrees of freedom, available to it by \eqref{eq:rankDef} not being satisfied, to greedily minimize the objective function $J(\xib, \ubs, \ybs)$. Given enough data (columns), it is established and realistic (see, e.g., \cite{Breschi2022new, Mattsson2024, KLAEDTKE2023} for a discussion) to make the following assumption.
\begin{assumption}\label{assum:fullRank}
    The data matrix $\Dbc$ has full row-rank.
\end{assumption}
Note that full row-rank of $\Dbc$ renders \eqref{eq:DPCeqConstr} meaningless without regularization, since there is an $\ab$ solving \eqref{eq:DPCeqConstr} for any arbitrary left-hand side trajectory. 

\subsection{Running numerical example}\label{sec:example}
We emphasize that this work does not propose a new DPC scheme; rather, it provides tools for analyzing the structure of the OCPs in existing schemes. Therefore, and in light of the following discussion in Section~\ref{sec:motivation} regarding point (i), we do not consider extensive closed-loop simulations to be the best demonstration of our results. Such simulations can be found in the works we reference when analyzing the respective schemes.
 Instead, we use a low-dimensional system as a running example to visualize the structure of the data matrix $\Dbc$ itself in Figure~\ref{fig:visualizeD}, and the structure of implicit predictors in Figure~\ref{fig:unconstrainedDPC} and~\ref{fig:terminalConstrainedDPC}. 
 Since using a past I/O sequence as a non-minimal state $\xib$ yields a minimum dimension of $(p+m)N_p \geq 2$,  we instead treat the example system in a state-space setting (see Remark~\ref{rem:Statespace}). The data-generating system is LTI with dimensions $n=m=1$ and parameters $(\Ab, \Bb)=(2, -0.5)$. To generate Figure~\ref{fig:visualizeD} (a), we drew 500 i.i.d. samples $\xb_0^{(i)}, \ubs^{(i)} \sim \Uc_{[-1,1]}$ and computed the subsequent state $\xbs^{(i)}$ via \eqref{eq:statespace_LTI}. For Figure~\ref{fig:visualizeD} (b), we added i.i.d. measurement noise $\Nc(0, 0.01)$ to each $\xb_0^{(i)}, \xbs^{(i)}$. To visualize the implicit predictors and optimal parametric solutions in Figure~\ref{fig:unconstrainedDPC}, we chose $N_f = 1$ and set up the control objective \eqref{eq:outputTrackingObj} with reference $\xbs_\text{ref}=0$, weight $\Qbc=1$, and quadratic input costs $J_\ubs(\xib, \ubs) = \ubs^2$. The data matrix $\Dbc$ is a block Hankel matrix (see, e.g., \cite{WILLEMS2005}), constructed from the input-state sequence
\begin{align*}
    \xb_{d,1} &= \begin{pmatrix}
        -0.1941 & 0.0048 & -0.2145 & -0.5427 & -1.3683
    \end{pmatrix}\\
    \ub_{d,1} &= \begin{pmatrix}
        -0.7859 & 0.4483 & \phantom{-}0.2274 & \phantom{-}0.5659
    \end{pmatrix},
\end{align*}
which was generated by drawing an initial state $\xb_0 \sim \Uc_{[-1,1]}$, applying the i.i.d. input sequence $\ub_d^{(i)}\sim \Uc_{[-1,1]}$ via \eqref{eq:statespace_LTI}, and then adding i.i.d. measurement noise $\Nc(0, 0.01)$ to the resulting state sequence.
To visualize the implicit predictors in Figure~\ref{fig:terminalConstrainedDPC}, we set $N_f = 2$, with reference $\xbs_\text{ref} = \begin{pmatrix} 0.5 & 0.5 \end{pmatrix}^\top$, weight $\Qbc = \Ib_2$, and arbitrary input costs, using a regularization weight of $\lambda = 1$. The data matrix $\Dbc$ is again a block Hankel matrix, constructed from the input-state sequence
\begin{align*}
    \xb_{d,2} &= \begin{pmatrix}
         \xb_{d,1} & -3.1384 & -5.9759
    \end{pmatrix}\\
    \ub_{d,2} &= \begin{pmatrix}
        \ub_{d,2} & \phantom{-}0.8037 & -0.6017
    \end{pmatrix},
\end{align*}
where the two additional samples are generated in the same way as before. These extra samples are needed to allow that the data matrix $\Dbc$ satisfies Assumption~\ref{assum:fullRank} for the increased prediction horizon $N_f = 2$.

\section{Motivation and setting} \label{sec:motivation}
Looking at recent DPC literature, we have observed the following two trends:
\begin{enumerate}
    \item[(i)] Regularization weights are typically tuned empirically, by evaluating closed-loop performance \cite{Dorfler2021, HUANG2021, Breschi2022new}.
    \item[(ii)] Heuristic interpretations of norm-based regularizations, e.g., $h(\ab) = \lambda\|\ab\|_2^2$ or $h(\ab) = \lambda\|\ab\|_1$, are (mostly unchanged) transferred to DPC \cite{Dorfler2021, Berberich2020stability}.
\end{enumerate}
We acknowledge that both observations are grounded in intuitive and practically valid heuristics. However, (i) only offers a quantitative indicator of control performance being improved/diminished by regularizations. It may obscure the underlying qualitative interactions between constraints, control objective, and regularization that led to this effect.
Some of the already established theoretical investigations in these qualitative effects include the interpretation of DPC with regularization as a convex relaxation of other (indirect) schemes \cite{Dorfler2021} or in the context of distributional robustness \cite{Coulson2019RegularizedDeePC}, and our work aims to add on to these.

Regarding (ii), such interpretations may be unspecific in the context of predictive control and might even be misleading. For example, it is claimed in \cite{Berberich2021_affineDPCexperimental} for regularization via $h(\ab) = \lambda\|\ab\|_2^2$ that too large $\lambda$ deteriorate tracking performance, since ``too small choices of the vector $\ab$ shift the input
and output to which the closed-loop converges towards
zero''\cite[Sec.~4]{Berberich2021_affineDPCexperimental}. While we agree that tracking performance is deteriorated, the inputs and outputs are generally not shifted towards zero by small $\ab$ for $\xib\neq\zerob$, since the chosen optimal $\ab^\ast$ must also satisfy $\Wb\ab^\ast = \xib$. Instead, they are shifted towards a (multistep) predictor $\hat \ybs(\xib, \ubs)$ and (multistep) controller $\hat\ubs(\xi)$ ``favored'' by the regularization $h(\ab)$. Both can be precisely characterized by our proposed tools, as demonstrated in Section~\ref{sec:trajectory-specific_quadratic} for linear DPC and Section~\ref{sec:affineDPC} for the affine DPC setting considered in \cite{Berberich2021_affineDPCexperimental}. 

The motivation of our work is to provide a structured analysis that explains the interactions between constraints, control objectives, and regularization in DPC. Although the current results focus on analyzing existing schemes rather than proposing new ones, they offer deeper insights into these interactions and reveal potential pitfalls. This creates a foundation for future improvements in DPC schemes. Furthermore, our analysis is intentionally agnostic to the specific class of the data-generating system. Rather than focusing on a particular system type, our setting is tailored to the structure of data matrices generated by them. This often reduces to Assumption~\ref{assum:fullRank}, which is typically (almost surely) satisfied for both non-deterministic LTI, and non-deterministic nonlinear systems. On the contrary, in practical cases, Assumption~\ref{assum:fullRank} typically only fails if too little data is used (i.e., $\Dbc$ is tall) or if the input data lacks sufficient excitation.
This generality comes at a  price. Without such assumptions on the true system, one cannot expect to evaluate how closely the analyzed predictions match its true behavior or how the optimal controls generated by DPC interact in closed-loop. Nonetheless, such assumptions can be readily added to tailor these results to a specific system class, incorporating prior system knowledge. Moreover, our findings can already help explain certain phenomena observed when applying DPC in closed-loop, such as those seen in \cite[Fig.~2]{Dorfler2021}, which we discuss in Section~\ref{sec:isolatingEffects}.

Our analysis relies on two key conceptual tools. One is given by the novel concept of \textit{implicit predictors} 
(depicted in  Figure~\ref{fig:implicitPredictor} and specified in Definition~\ref{def:implicit_predictor} further below), which aims to describe the predictive behavior of OCPs that may not have an explicitly enforced prediction model. The introduction of implicit predictors in \cite{KLAEDTKE2023} has also led to a parametric characterization of regularization costs in terms of their \textit{trajectory-specific effects} (formalized in Definition~\ref{def:trajectorySpecific}). Since then, similar analyses have appeared in \cite{Mattsson2024, Klaedtke2024FCS_DPC, klaedtke2024unifyingPreprint, Breschi2023regularizationImpact}, offering useful generalizations and intuitive interpretations of the underlying structures. Due to its independent value beyond implicit predictors, we treat this as a separate, second tool.

We first introduce the trajectory-specific effect of regularizations in Section~\ref{sec:trajectory-specific}, since it facilitates the following introduction of implicit predictors in Section~\ref{sec:implicitPredictors}. As an exemplary demonstration of these tools, we summarize results from \cite{KLAEDTKE2023, Klaedtke2024FCS_DPC, klaedtke2024unifyingPreprint} for their use in DPC with quadratic regularization in Section~\ref{sec:trajectory-specific_quadratic}, \ref{sec:isolatingEffects}, and Section~\ref{sec:implicitPredictors_quadratic}.
 We then extend these results in Section~\ref{sec:extensions} by analyzing the effects of common modifications to the DPC problem. These analyses cover the extension to affine DPC in Section~\ref{sec:affineDPC}, the inclusion of an offset in the regularization in Section~\ref{sec:offset}, the inclusion of slack variables in Section~\ref{sec:Slack}, and the inclusion of additional (terminal) equality constraints in Section~\ref{sec:terminalConstraints}. While not technically a modification to DPC, we also give two short but very general results on (recursive) feasibility in DPC in Section~\ref{sec:feasibility}.

\section{Trajectory-specific effect of regularization}\label{sec:trajectory-specific}

In the ideal deterministic LTI setting without regularization, the variable $\ab$ is only used in \eqref{eq:DPCeqConstr} as an expression for the image representation introduced in \eqref{eq:imageRep0}, i.e.,
\begin{equation}
    \begin{pmatrix}
     \xib \\ \ubs \\ \ybs 
\end{pmatrix}
    \in
    \image{\Dbc} \quad \iff \quad \exists \ab\; \text{such that}\; 
    \begin{pmatrix}
     \xib \\ \ubs \\ \ybs 
\end{pmatrix} = \Dbc\ab. \label{eq:imageRep}
\end{equation}
However, adding a regularization $h(\ab)$ introduces another meaning to $\ab$, which is not based on the image representation and the underlying behavioral system theory but on heuristics. Intuitively, $h(\ab)$ adds a price tag to every $\ab$, which is also transferred to the trajectory tuple $(\xi, \ubs, \ybs)$ generated by $\Dbc \ab$. Note that, if $\Dbc$ is wide, there might be multiple $\ab$ generating the same trajectory and thus multiple price tags associated with it. However, since we are optimizing over $\ab$ in \eqref{eq:DPC}, it is natural to only consider the lowest cost associated with any trajectory. This is formally captured by the following definition. 
\begin{definition}\label{def:trajectorySpecific}
    We call the solution $h^\ast(\xib, \ubs, \ybs)$ to the optimization problem
\begin{subequations}\label{eq:OP_trajectorySpecific}
\begin{align} 
h^\ast(\xib, \ubs, \ybs) &:= h(\ab^\ast(\xib, \ubs, \ybs)) \nonumber\\
&=  \min_{\ab} 
  h(\ab) \quad  \text{s.t.} \label{eq:OP_trajectorySpecific_cost}\\
& \quad \quad  \begin{pmatrix}
     \xib \\ \ubs \\ \ybs 
\end{pmatrix} = \begin{pmatrix}
    \Wb \\ \Ub \\ \Yb 
\end{pmatrix}\ab \label{eq:OP_trajectorySpecific_constr}
\end{align}
\end{subequations}
 the \textit{trajectory-specific effect} of the regularization $h(\ab)$ given the data $\Dbc$.
\end{definition}
 Importantly, note that $(\xib, \ubs, \ybs)$ appear as parameters in $\eqref{eq:OP_trajectorySpecific}$ and not as optimization variables. Therefore, additional constraints such as \eqref{eq:DPCsetConstr} are irrelevant. That is, $h^\ast(\xib, \ubs, \ybs)$ is valid for all $(\xib, \ubs, \ybs)$ satisfying \eqref{eq:imageRep}, and therefore also for those, which additionally need to satisfy \eqref{eq:DPCsetConstr}. 
The relevance of $h^\ast(\xib, \ubs, \ybs)$ to the DPC problem comes from the fact that \eqref{eq:OP_trajectorySpecific} naturally appears as an inner optimization problem in \eqref{eq:DPC}. That is, \eqref{eq:DPC} is equivalent to
\begin{subequations}
\label{eq:DPC_without_a}
\begin{align}
\min_{\ubs,\ybs} 
 J(\xib, \ubs, &\ybs) + h^\ast(\xib, \ubs, \ybs) \label{eq:DPC_without_a_cost} \\
\text{s.t.} \quad \quad  \begin{pmatrix}
     \xib \\ \ubs \\ \ybs 
\end{pmatrix} & \in
    \image{\Dbc}, \label{eq:DPC_without_a_imageConstr}\\
\left(\ubs, \ybs \right) &\in \Uc \times \Yc.  \label{eq:DPC_without_a_setConstr}
\end{align}
\end{subequations}
Note that we have deliberately replaced \eqref{eq:DPCeqConstr} via \eqref{eq:imageRep} to highlight the fact that $\ab$ can be fully eliminated, since it is just an auxiliary variable, after all. That is, the image representation \eqref{eq:imageRep} acts just as before, but the additional heuristic costs introduced by $h(\ab)$ (with which we started this section) are now fully explained by their trajectory-specific effect $h^\ast(\xib, \ubs, \ybs)$. This allows for much more intuitive interpretations, which we demonstrate by summarizing results from \cite{KLAEDTKE2023, Klaedtke2024FCS_DPC, klaedtke2024unifyingPreprint} on the trajectory-specific effect of quadratic regularization $h(\ab) = \lambda \|\ab\|_2^2$ in Section~\ref{sec:trajectory-specific_quadratic}, and its projection-based variant $h(\ab) = \lambda \|\Pib_\perp\ab\|_2^2$ in Section~\ref{sec:isolatingEffects}, where we also discuss its link to $\gammab$-DDPC \cite{Breschi2022new}. One may notice that, due to its popularity and its conformity for analytic solutions, all results in this paper focus on (projection-based) quadratic regularization. However, we want to emphasize the generality of Definition~\ref{def:trajectorySpecific}, which is suited for \textit{any} choice of regularization $h(\ab)$. In particular, similar analyses of trajectory-specific effects for 1-norm $h(\ab)=\lambda \|\ab\|_1$ \cite{Coulson2019DeePC} or general p-norm $h(\ab)=\lambda \|\ab\|_p$ \cite{Dorfler2021} regularizations should be equally insightful, but are still an open problem.

\subsection{Trajectory-specific effect of quadratic regularization}\label{sec:trajectory-specific_quadratic}

\begin{figure*}
    \centering
    \includegraphics[trim=1.6cm 12.35cm 0.9cm 12.42cm,clip=true, scale=0.9]{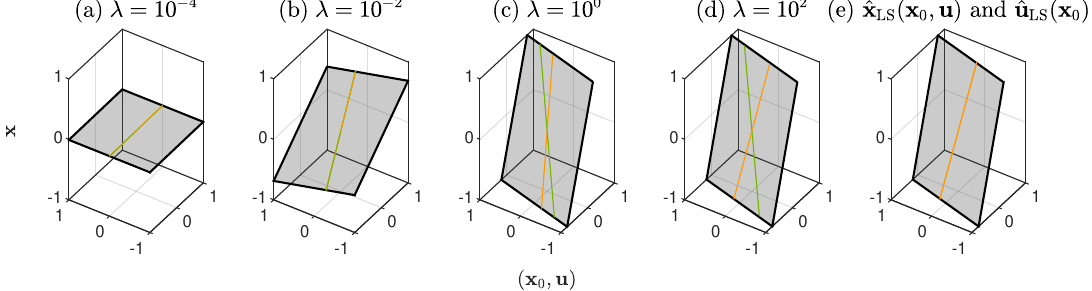}
        \caption{
        Implicit predictor, optimal parametric solutions, and least-squares mappings for the DPC problem discussed in Section 2.3. (a-d) The optimal parametric DPC solutions $(\xb_0, \ubs^\ast(\xb_0), \xbs^\ast(\xb_0))$ for the different regularizations $h(\ab) = \lambda \|\ab\|_2^2$ (orange) and $h(\ab) = \lambda \|\Pib_\perp\ab\|_2^2$ (green) naturally evolve on the implicit predictor $\hat\xbs_\text{DPC}(\xb_0, \ubs)$ (gray). (e) Visualization of the least-square solutions $\hat\xbs_\text{LS}(\xb_0, \ubs)$ (gray) and $\hat\ubs_\text{LS}(\xb_0)$. The latter is shown via the tuple $(\xb_0, \hat\ubs_\text{LS}(\xb_0), \hat\xbs_\text{LS}(\xb_0, \hat\ubs_\text{LS}(\xb_0)))$ (orange).}
        \label{fig:unconstrainedDPC}
\end{figure*}

It was first shown in \cite{KLAEDTKE2023} with additional details in \cite[Prop.~1]{klaedtke2024unifyingPreprint} that, under Assumption~\ref{assum:fullRank}, the trajectory-specific effect of quadratic regularization $h(\ab) = \lambda \|\ab\|_2^2$ is given by
\begin{subequations}\label{eq:2NormCost}
    \begin{align}
        \lambda\|\ab^\ast(\xib, \ubs, \ybs)\|_2^2 &= \lambda \|\ybs-\hat\ybs_\text{LS}(\xib, \ubs)\|_{\Qbc_\text{reg}}^2 \label{eq:2NormCostTerm1}\\
    &\qquad + \lambda \|\ubs-\hat\ubs_\text{LS} (\xib)\|_{\Rbc_\text{reg}}^2 \label{eq:2NormCostTerm2}\\
    & \qquad+ \lambda\|\xib\|_{\left(\Wb\Wb^\top\right)^{-1}}^2. \label{eq:2NormCostTerm3}
\end{align}
\end{subequations}
Here, $\hat\ybs_\text{LS}(\xib, \ubs)$ and $\hat\ubs_\text{LS}(\xib)$ are the (multistep) predictor/controller ``favored'' by the regularization, as initially mentioned in Section~\ref{sec:motivation}. The regularization pushes predicted $\ybs$ and $\ubs$ towards them, which is visualized in Figure~\ref{fig:unconstrainedDPC} for the low-dimensional state-space example discussed in Section~\ref{sec:example}. By parametrically solving \eqref{eq:OP_trajectorySpecific} and decomposing block-matrix expressions (see \cite{KLAEDTKE2023} for details), one can show that both are given by linear mappings, which can be represented as
\begin{align}
    \hat\ybs_\text{LS}(\xib, \ubs) &:= \Gb_\text{LS} \zb\\
    \hat\ubs_\text{LS}(\xib) &:= \Kb_\text{LS} \xib
\end{align}
via the least-squares solutions
\begin{subequations}\label{eq:LSmappings}
    \begin{align}
    \Gb_\text{LS} &:= \argmin_\Gb \|\Yb - \Gb \Zb\|_\mathrm{F}^2 = \Yb \Zb^+ \label{eq:G_LS}\\
    \Kb_\text{LS} &:= \argmin_\Kb \|\Ub - \Kb \Wb\|_\mathrm{F}^2 = \Ub \Wb^+,
\end{align}
\end{subequations}
where $\|.\|_\mathrm{F}$ denotes the Frobenius norm.
Note that $\hat\ybs_\text{LS}(\xib, \ubs)$ is equivalent to the subspace predictor used as an equality constraint in subspace predictive control (SPC, \cite{FAVOREEL1999}). The involved weighing matrices are given by 
\begin{equation}
    \Qbc_\text{reg} := \left(\dYb\dYb^\top\right)^{-1} \;\; \text{and} \;\;  \Rbc_\text{reg} := \left(\dUb\dUb^\top\right)^{-1}, \label{eq:regWeights}
\end{equation}
where
$$
    \dYb := \Yb - \Gb_\text{LS} \Zb \quad \text{and} \quad 
    \dUb := \Ub - \Kb_\text{LS} \Wb
$$
are the residual matrices associated with the least-square problems \eqref{eq:LSmappings}, and the inverses exist under Assumption~\ref{assum:fullRank}. Therefore, $\Qbc_\text{reg}$ and $\Rbc_\text{reg}$ can be interpreted as inverses of scaled (because they are not normalized w.r.t. the amount of data columns $\ell$) empirical second moment matrices for the output prediction error $\dYb$ and ``input prediction error'' $\dUb$ of the least-squares solutions $\Gb_\text{LS}, \Kb_\text{LS}$ based on the data $\Dbc$. Hence, quadratic regularization $h(\ab) = \lambda \|\ab\|_2^2$ pushes $\ybs$ and $\ubs$ towards the least-squares estimates $\hat\ybs_\text{LS}(\xib, \ubs) $ and $\hat\ubs_\text{LS}(\xib) $. It does so more (less) harshly in directions, where the latter are believed to be more (less) accurate, based on the available data. 
The same second moment (or covariance)-based interpretation appears in \cite{Mattsson2024}, with the distinction that the state-input dependent part in \cite[Eq.~(21)]{Mattsson2024} is not yet decomposed into the two terms \eqref{eq:2NormCostTerm2} and \eqref{eq:2NormCostTerm3}, similar to \cite{KLAEDTKE2023}. However, \cite{Mattsson2024} also shows that the results naturally generalize to rank-deficient data matrices, i.e., $\Dbc$ not satisfying Assumption~\ref{assum:fullRank}, by simply using the Moore-Penrose inverse in \eqref{eq:regWeights}. In that case, deviations $\dyb = \ybs - \hat\ybs_\text{LS}(\xib, \ubs)$ are not penalized in singular directions of $\dYb \dYb^\top$, i.e., directions in which no deviation from $\hat\ybs_\text{LS}(\xib, \ubs)$ has been observed in the data matrix $\Dbc$. However, due to \eqref{eq:imageRep}, one can show  that $\ybs$ may not deviate at all in these directions. That is, in unexplored directions of the state-input-output space, the penalization of deviations in \eqref{eq:2NormCost} is replaced with hard constraints, since deviations from it are not possible in $\image{\Dbc}$. The same considerations apply for deviations $\dub = \ubs - \hat\ubs_\text{LS}(\xib)$ if $\Zb$ does not have full row-rank.

When discussing the role of these cost terms, first note that the last cost term \eqref{eq:2NormCostTerm3} is irrelevant to the OCP, since $\xib$ is a parameter determined in closed-loop and not an optimization variable.
Regarding the usefulness of \eqref{eq:2NormCostTerm1}, we believe that $\ybs$ being pushed towards the least-squares (multistep) predictor $\hat\ybs_\text{LS}(\xib, \ubs)$ in \eqref{eq:2NormCostTerm1} is quite intuitive. However, while $\hat\ybs_\text{LS}(\xib, \ubs)$ is conceptually not a bad choice for a predictor, it should be noted that (without further modifications such as in \cite{sader2023causalityinformed}) it is typically neither causal nor time-invariant. Furthermore, although it is conceptually nice to include additional statistical information via $\Qbc_\text{reg}$ and $ \Rbc_\text{reg}$, this information is utilized very greedily in DPC. That is, since the optimal solution is a trade-off between control objective and regularization cost, the DPC cost objective \eqref{eq:DPCcost} promotes deviations from $\hat\ybs_\text{LS}(\xib, \ubs)$ primarily in uncertain directions in order to achieve lower costs for the control objective $J(\xib, \ubs, \ybs)$. We believe that it remains an interesting avenue for future research to explore under what conditions penalizing predictions $\ybs$ towards a (potentially well- or ill-suited) predictor $\hat\ybs(\xib, \ubs)$, as in \eqref{eq:2NormCostTerm1}, outperforms enforcing it as an equality constraint $\ybs = \hat\ybs(\xib, \ubs)$.
While the intuition behind penalizing the input $\ubs$ towards $\hat\ubs_\text{LS}(\xib)$ in the same fashion may not be immediately clear, it was shown in the $\gammab$-DDPC setting that a cost term equivalent to \eqref{eq:2NormCostTerm2} (proven in \cite{KLAEDTKE2023}) can indeed be utilized to increase DPC performance \cite{BRESCHI2023uncertaintyAware}. Furthermore, a functionally similar input penalty appears in another framework focused on minimizing the final control error (FCE) \cite{chiuso2024FCE}. Intuitively, the performance improvement can be explained by the term \eqref{eq:2NormCostTerm2} pushing the chosen input $\ubs$ towards the best-explored region of the state-input-space, i.e., where the most confident predictions can be made based on the available data. For a more precise quantification of the relation between prediction error variance and $\|\gammab_2\|_2^2$, we refer to \cite{BRESCHI2023uncertaintyAware}.

Finally, we want to highlight the discrepancy between \eqref{eq:2NormCostTerm1} and \eqref{eq:2NormCostTerm2} in terms of proper tuning for $\lambda$. The output-related cost term \eqref{eq:2NormCostTerm1} needs a large weight $\lambda$ because, without the previously discussed rank deficiency \eqref{eq:rankDef}, it is the sole factor keeping output predictions $\ybs$ from being greedily and unrealistically (i.e., without considering the data in $\Dbc$) pushed toward the minimum of the control objective $J(\xib, \ubs, \ybs)$. This effect is visualized in Figure~\ref{fig:unconstrainedDPC}(a). On the other hand, the input-related cost term \eqref{eq:2NormCostTerm2} may not have a large weight $\lambda$, since this would favor always sticking to the ``best-explored'' choice $\hat\ubs_\text{LS}(\xib)$ over choosing an input sequence $\ubs$ that benefits the control objective $J(\xib, \ubs, \ybs)$, which is visualized in Figure~\ref{fig:unconstrainedDPC}(d-e). These considerations highlight the need for isolation of the individual cost terms.  In the following section, we present two modified DPC schemes in which, as revealed through an analysis of the trajectory-specific effect of regularization, this isolation of regularization cost terms naturally occurs.

\subsection{Isolating trajectory-specific effects via projections or $\gamma$-DDPC}\label{sec:isolatingEffects}

In \cite{Dorfler2021}, the orthogonal projection matrices
$$
    \Pib := \Zb^+ \Zb \quad \text{and} \quad \Pib_\perp := \Ib - \Pib
$$
were introduced to yield a regularization $h(\ab) = \lambda \| \Pib_\perp \ab\|_2^2$ for which the closed-loop behavior is consistent with the subspace predictor as $\lambda \to \infty$. While it is noted that norm-based regularizers such as $h(\ab) = \lambda \| \ab\|_2^2$ ``are not consistent and bias the optimal solution'' \cite[Sec.~IV.C]{Dorfler2021}, the nature of this bias is not further explored. In \cite{KLAEDTKE2023}, this nature is uncovered by showing that (under Assumption~\ref{assum:fullRank}) the trajectory-specific effect of projection-based quadratic regularization $h(\ab) = \lambda \| \Pib_\perp \ab\|_2^2$ is given by 
\begin{equation}\label{eq:proj2NormCost}
    \lambda\|\Pib_\perp\ab^\ast(\xib, \ubs, \ybs)\|_2^2 = \lambda \|\ybs-\hat\ybs_\text{LS}(\xib, \ubs)\|_{\Qbc_\text{reg}}^2.
\end{equation}
Since $\|\ab\|_2^2 = \| \Pib \ab\|_2^2 + \| \Pib_\perp \ab\|_2^2$, we also immediately see that the remaining cost terms \eqref{eq:2NormCostTerm2} and \eqref{eq:2NormCostTerm3} are associated with $\Pib$. That is, (under Assumption~\ref{assum:fullRank}) the trajectory-specific effect of projection-based quadratic regularization $h(\ab) = \lambda \| \Pib \ab\|_2^2$ is given by 
$$
\lambda\|\Pib\ab^\ast(\xib, \ubs, \ybs)\|_2^2 = \lambda \|\ubs-\hat\ubs_\text{LS} (\xib)\|_{\Rbc_\text{reg}}^2 + \lambda\|\xib\|_{\left(\Wb\Wb^\top\right)^{-1}}^2.
$$
This separation explains the observations in \cite[Fig.~2]{Dorfler2021}. There, the performance of $h(\ab) = \lambda \| \ab\|_2^2$ deteriorates for $\lambda \to \infty$ since the chosen input $\ubs$ is aggressively pushed towards $\hat\ubs_\text{LS} (\xib)$, while neglecting the control objective $J(\xib, \ubs, \ybs)$. In contrast, the projection-based regularization $h(\ab) = \lambda \| \Pib_{\perp} \ab\|_2^2$ avoids this problem by fully dropping the input-related term \eqref{eq:2NormCostTerm2}. However, the ``closed-loop view'' (discussed in Section~\ref{sec:motivation}) in \cite[Fig.~2]{Dorfler2021} hides the fact that two functionally very different cost terms \eqref{eq:2NormCostTerm1} and \eqref{eq:2NormCostTerm2} are responsible for the observed performance. Furthermore, \cite{Dorfler2021} only considers the two options of either using both \eqref{eq:2NormCostTerm1} and \eqref{eq:2NormCostTerm2} with the same weight $\lambda$ via $h(\ab) = \lambda \| \ab\|_2^2$, or only using \eqref{eq:2NormCostTerm1} via $h(\ab) = \lambda \| \Pib_\perp \ab\|_2^2$. However, an individually weighted mix of the two terms may provide the best of both worlds, which is coincidentally explored in the regularization schemes proposed for $\gammab$-DDPC \cite{Breschi2022new}.

In $\gammab$-DDPC, the constraint \eqref{eq:DPCeqConstr} and variable $\ab$ are replaced via LQ decomposition as
$$
\begin{pmatrix}
        \Wb \\
        \Ub \\
        \Yb
    \end{pmatrix} 
    = 
    \begin{pmatrix}
        \Lb_{11} & \zerob & \zerob & \zerob \\
        \Lb_{21} & \Lb_{22} & \zerob & \zerob \\
        \Lb_{31} & \Lb_{32} & \Lb_{33}  & \zerob
    \end{pmatrix}
    \begin{pmatrix}
        \Qb_1 \\
        \Qb_2 \\
        \Qb_3 \\
        \Qb_4
    \end{pmatrix}, 
     \gammab := \begin{pmatrix}
        \gammab_1 \\ 
        \gammab_2 \\
        \gammab_3 \\
        \gammab_4
    \end{pmatrix} = \begin{pmatrix}
        \Qb_1 \\
        \Qb_2 \\
        \Qb_3 \\
        \Qb_4
    \end{pmatrix} \ab, 
$$
where the diagonal blocks $\Lb_{ii}$ for $i \in \{1, 2, 3\}$ are non-singular (under Assumption~\ref{assum:fullRank}) and the matrices $\Qb_i$ have orthonormal rows. Furthermore, $\Qb_4$ and $\gammab_4$ are typically omitted, since they do not affect the generated trajectory. The idea behind $\gammab$-DDPC can be summarized as re-parameterizing the OCP with a lower dimensional variable $\gammab$, and decoupling the matching of the initial condition $\xib$, since $\gammab_1 =\Lb_{11}^{-1} \xib$ is uniquely determined \cite{Breschi2022new}. The proposed regularization strategies for $\gammab$-DDPC are based on a mix of quadratic regularization $\tilde h(\gammab) = \lambda_2\|\gammab_2\|_2^2+\lambda_3\|\gammab_3\|_2^2$, or constraining $\gammab_3 = \zerob$ and using only $\tilde h(\gammab) = \lambda_2\|\gammab_2\|_2^2$ \cite{BRESCHI2023uncertaintyAware, Breschi2023regularizationImpact}. The connection between this regularization of $\gammab$-variables and projection-based regularization of $\ab$ was revealed in \cite{klaedtke2024unifyingPreprint} by analyzing their trajectory-specific effect. 
\begin{corollary}[\cite{klaedtke2024unifyingPreprint}]\label{corr:gammaDDPCcost}
        Under Assumption~\ref{assum:fullRank}, regularization of the $\gammab$-variables can be equivalently expressed by the trajectory-specific effect
        \begin{align*}
            \|\gammab_1\|_2^2 &= \|\xib\|_{\left(\Wb\Wb^\top\right)^{-1}}^2,\\
            \|\gammab_2\|_2^2 &= \|\ubs-\hat\ubs_\mathrm{LS}(\xib)\|_{\Rbc_\mathrm{reg}}^2,\\
            \|\gammab_3\|_2^2 &= \|\ybs-\hat\ybs_\mathrm{LS}(\xib, \ubs)\|_{\Qbc_\mathrm{reg}}^2.
        \end{align*}
    \end{corollary}
We note that the difference in notation w.r.t. to Definition~\ref{def:trajectorySpecific} only comes from the fact that $(\gammab_1, \gammab_2, \gammab_3)$ are uniquely determined by $(\xib, \ubs, \ybs)$, and therefore $\gammab_i^\ast(\xib, \ubs, \ybs) = \gammab_i$. From Corollary~\ref{corr:gammaDDPCcost}, we can see that both the projection-based regularization and the regularization of $\gammab_i$ isolate specific terms of the trajectory-specific effect of standard quadratic regularization. Their relation can be summarized as follows
\begin{align*}
      &\overbrace{\underbrace{\|\ybs-\hat\ybs_\text{LS}(\xib, \ubs)\|_{\Qbc_\text{reg}}^2}_{\|\gammab_3\|_2^2}}^{\|\Pib_\perp\ab^\ast(\xib, \ubs, \ybs)\|_2^2}  +  \overbrace{\underbrace{\|\ubs-\hat\ubs_\text{LS} (\xib)\|_{\Rbc_\text{reg}}^2}_{\|\gammab_2\|_2^2} + \underbrace{\|\xib\|_{\left(\Wb\Wb^\top\right)^{-1}}^2 }_{\|\gammab_1\|_2^2}}^{\|\Pib\ab^\ast(\xib, \ubs, \ybs)\|_2^2} \\
      &= \|\ab^\ast(\xib, \ubs, \ybs)\|_2^2.
\end{align*}
As a guideline for practitioners, we advocate to use a mixed regularization $h(\ab) = \lambda_2 \|\Pib \ab\|_2^2+\lambda_3 \|\Pib_\perp \ab\|_2^2$ (or, equivalently in the $\gammab$-DDPC framework $\tilde h(\gammab) = \lambda_2 \|\gammab_2\|_2^2+\lambda_3 \|\gammab_3\|_2^2$). This approach generalizes both the quadratic regularization $h(\ab) = \lambda \|\ab\|_2^2$, and the purely projection based $h(\ab) = \lambda \|\Pib_\perp\ab\|_2^2$ by choosing $\lambda = \lambda_2=\lambda_3$ or $\lambda = \lambda_3, \lambda_2=0$, respectively. To avoid output predictions blindly following the control objective, $\lambda_3$ should be chosen very large, and (except for potential numerical solver instabilities) there are no adverse effects to be expected even for $\lambda_3 \to \infty$. On the contrary, $\lambda_2$ should be chosen very carefully to avoid the inputs blindly following the least-squares controller $\hat\ubs_\text{LS}(\xib)$. While $\lambda_2=0$ (recovering the projecion-based regularization $h(\ab)=\lambda_3\|\Pib_\perp\ab\|_2^2$) can be considered a safe choice in this regard, it also foregoes the potential performance improvements associated with proper tuning of this term. Here, we refer to \cite{BRESCHI2023uncertaintyAware} for more specific tuning guidelines. Finally, as also observed in \cite{Mattsson2024}, adding or removing trajectory columns from $\Dbc$ after the tuning requires a re-scaling of all regularization weights
$$
    \lambda_{i,\text{new}} = \frac{\ell_\text{new}}{\ell_\text{old}} \lambda_{i,\text{old}}
$$
to retain a quantitatively similar effect, since $\Qbc_\text{reg}, \Rbc_\text{reg}$ are not normalized w.r.t. the number of trajectory columns.

\begin{remark}
    Technically, the data matrix $\Dbc$ is already normalized w.r.t. $\sqrt{\ell}$ in the $\gammab$-DDPC framework, such that the aforementioned re-scaling is not necessary. For clarity of presentation and direct comparison to the projection-based regularization, we have omitted this normalization. Note that all considerations remain valid with(out) normalization, since it does not change the image representation \eqref{eq:imageRep} and the same effect in terms of quadratic regularization can be achieved by rescaling weights $\lambda_i$ with $\ell$.
\end{remark}

\section{Implicit predictors in regularized DPC}\label{sec:implicitPredictors}

Although the absence of the rank deficiency \eqref{eq:rankDef} allows for non-unique output predictions, and Assumption~\ref{assum:fullRank} even allows for \textit{any} triple $(\xib, \ubs, \ybs)$ to be generated from $\Dbc$ via linear combinations, one can still observe certain patterns in the optimal solutions $(\ybs^\ast(\xib), \ubs^\ast(\xib), \xib)$ to \eqref{eq:DPC}. In particular, it seems that the output predictions implicitly follow a (multistep) predictor mapping $\ybs^\ast = \hat\ybs(\xib, \ubs^\ast(\xib))$, even though such a mapping is not explicitly enforced as an equality constraint in \eqref{eq:DPC}. These considerations led to the following notion of implicit predictors introduced in \cite{KLAEDTKE2023}.
\begin{definition}[\cite{KLAEDTKE2023}]\label{def:implicit_predictor}
    We call $\hat\ybs(\xib,\ubs)$ an \textit{implicit predictor} for an OCP if including the constraint $\ybs = \hat\ybs(\xib, \ubs)$ does not alter the (set of) minimizers $(\ubs^\ast, \ybs^\ast)$ and the optimal value.    
\end{definition}
Hence, an implicit predictor $\hat\ybs_\text{DPC}(\xib, \ubs)$ for the DPC problem \eqref{eq:DPC} can be interpreted as a model that generates the output predictions $\ybs$ based on $(\xib, \ubs)$ consistent with the solutions to \eqref{eq:DPC}, even though such a model is never explicitly enforced. It is meant as a descriptive object (not prescriptive, since it is not enforced in \eqref{eq:DPC}), which can be studied to explain the predictive behavior of DPC schemes and make improvements, accordingly. The interaction of this interpretation with the direct and indirect data-driven control paradigms is visualized in Figure~\ref{fig:implicitPredictor}.
While Definition~\ref{def:implicit_predictor} is more conceptional than constructive, a valid implicit predictor for the DPC problem \eqref{eq:DPC} can be constructed by solving the optimization problem
\begin{subequations}
\label{eq:implicitPredictorOP}
\begin{align}
\hat\ybs_\text{DPC}(\xib, \ubs) = \argmin_\ybs \min_{\ab} 
 J(\xib, \ubs, &\ybs) + h(\ab) \label{eq:implicitPredictorOPcost} \\
\text{s.t.} \quad \quad  \begin{pmatrix}
     \xib \\ \ubs \\ \ybs 
\end{pmatrix} &= \begin{pmatrix}
    \Wb \\ \Ub \\ \Yb 
\end{pmatrix}\ab, \label{eq:implicitPredictorOPeqConstr}\\
 \ybs  &\in \Yc.  \label{eq:implicitPredictorOPsetConstr}
\end{align}
\end{subequations}
In contrast to \eqref{eq:OP_trajectorySpecific}, we now treat $(\xib, \ubs)$ as parameters and optimize over $(\ab, \ybs)$. Hence, additional set constraints $\ubs\in \Uc$ can be dropped while $\ybs \in \Yc$ still need to be considered. However, similar to \eqref{eq:OP_trajectorySpecific}, note that \eqref{eq:implicitPredictorOP} is an inner optimization problem to \eqref{eq:DPC} and thus its parametric solution $\hat\ybs_\text{DPC}(\xib, \ubs)$ naturally satisfies Definition~\ref{def:implicit_predictor}.

In the presence of the rank deficiency \eqref{eq:rankDef}, $\hat\ybs_\text{DPC}(\xib, \ubs)$ is fully determined by the image representation \eqref{eq:implicitPredictorOPeqConstr}. Furthermore, it reveals the unique linear predictor mapping implicitly within the data matrix $\Dbc$, which can be specified as $\hat\ybs_\text{DPC}(\xib, \ubs) = \hat\ybs_\text{LS}(\xib, \ubs)$. Note that this is precisely the setting and results considered in \cite{Fiedler2021}. That is, in the deterministic LTI case \cite[Thm.~1]{Fiedler2021}, the rank deficiency \eqref{eq:rankDef} naturally occurs, as discussed in Section~\ref{sec:fundamentals}. 
Notably, in the non-deterministic setting,  \cite[Thm.~2]{Fiedler2021} considers a data matrix $\Dbc$ which has so few columns $\ell$ that the rank deficiency \eqref{eq:rankDef} is also satisfied, i.e., 
\begin{equation}
    \ell = \rank{\Zb} \quad \implies \quad \rank{\Dbc} = \rank{\Zb}. \label{eq:smallEll}
\end{equation}
However, in these two settings, the predictor $\hat\ybs_\text{LS}(\xib, \ubs)$ somewhat loses its meaning as a least squares estimate \eqref{eq:G_LS}, as it consistently yields zero residual ($\dYb = \zerob$). That is, in the presence of rank deficiency \eqref{eq:rankDef}, the linear system of equations \eqref{eq:imageRep} can be solved algebraically for an exact and unique linear mapping $\hat\ybs(\xib, \ubs)$, and using the least squares solution is merely one way of doing so. In other words, \cite{Fiedler2021} is often cited when arguing for equivalence between SPC and DPC. However, this equivalence only occurs if there is an exact linear output predictor mapping within the data $\Dbc$, since both methods manage to find it. Any alternative method that successfully identifies this exact linear mapping and employs it for predictive control would also be considered equivalent. 

In the absence of the rank-deficiency \eqref{eq:rankDef}, i.e., in a realistic (non-deterministic) setting with more data columns than in \eqref{eq:smallEll}, the additional degrees of freedom lead to output predictions deviating from $\hat\ybs_\text{LS}(\xib, \ubs)$. Although it was shown in \cite{Dorfler2021} that the optimal solution of DPC with projection-based quadratic regularization $h(\ab) = \lambda\|\Pib_\perp\ab\|_2^2$ is consistent with SPC for $\lambda \to \infty$ (and therefore $\hat\ybs_\text{DPC}(\xib, \ubs) = \hat\ybs_\text{LS}(\xib, \ubs)$ in that case), a similar characterization of predictive behavior for $h(\ab) = \lambda\|\ab\|_2^2$ or finite $\lambda$ is not discussed 
in \cite{Dorfler2021}. Precisely this kind of characterization of predictive behavior for finite $\lambda$ is possible using the concept of implicit predictors and results for (projection-based) quadratic regularization are given in \cite{KLAEDTKE2023, Klaedtke2024FCS_DPC}, which will be summarized in the following section.
Although we focus on DPC with these two kinds of quadratic regularization, we want to emphasize the generality of Definition~\ref{def:implicit_predictor}, which is suited for any kind of predictive control scheme.

\subsection{An implicit predictor for DPC with quadratic regularization}\label{sec:implicitPredictors_quadratic}

In the following, we assume that the control objective is a quadratic output-tracking formulation
\begin{equation}\label{eq:outputTrackingObj}
    J(\xib, \ubs, \ybs) = \|\ybs-\ybs_\text{ref}\|_\Qbc^2 + J_\ubs(\xib, \ubs)
\end{equation}
with reference $\ybs_\text{ref}$, positive semidefinite weighing matrix $\Qbc$, and arbitrary input control objective $J_\ubs(\xib, \ubs)$. We further assume that no additional output constraints are present, i.e., $\Yc = \R^{p N_f}$. For some general observations on the effect of output constraints, see \cite[Sec.~III.C]{KLAEDTKE2023}, and specific results regarding (terminal) equality constraints and general feasibility are given in Section~\ref{sec:terminalConstraints} and Section~\ref{sec:feasibility}, respectively. In \cite[Thm~3, 4]{KLAEDTKE2023}, results for an implicit predictor with reference $\ybs_\text{ref}=\zerob$ are given, which were later extended to possibly nonzero reference in \cite[Thm.~2]{Klaedtke2024FCS_DPC}. Under Assumption~\ref{assum:fullRank}, solving \eqref{eq:implicitPredictorOP} comes down to solving the unconstrained optimization problem
$$
        \hat\ybs_\text{DPC}(\xib, \ubs)=\arg\min_{\ybs} 
     \|\ybs-\ybs_\text{ref}\|_{\Qbc}^2  + h^\ast(\xib, \ubs, \ybs).
$$
Furthermore, since the trajectory-specific cost of $h(\ab) = \lambda \|\ab\|_2^2$ and $h(\ab) = \lambda \|\Pib_\perp\ab\|_2^2$ only differ in terms independent of $\ybs$ (see \eqref{eq:2NormCost} and \eqref{eq:proj2NormCost}), it turns out that the predictive behavior of both DPC schemes can be characterized by the same implicit predictor
\begin{align}
 \label{eq:DPC_predictor}
        \hat\ybs_\text{DPC}(\xib, \ubs)=&
        \left(\lambda \Qbc_\text{reg}+\Qbc\right)^{-1} \lambda \, \Qbc_\text{reg}\, \hat\ybs_\text{LS}(\xib,\ubs)\\
        &\;+\left(\lambda \Qbc_\text{reg}+\Qbc\right)^{-1} \Qbc \, \ybs_\text{ref}. \nonumber
\end{align}
Note that the involved inverse exists because $\Qbc_\text{reg}$ is positive definite (under Assumption~\ref{assum:fullRank}) and $\Qbc$ is positive semidefinite.
Structurally, \eqref{eq:DPC_predictor} can be seen as a \mbox{(matrix-)weighed} sum, shifting output predictions towards $\ybs_\text{ref}$ and $\hat\ybs_\text{LS}(\xib, \ubs)$, depending on the weights $\Qbc$ and $\lambda\Qbc_\text{reg}$, respectively. This structure is visualized in Figure~\ref{fig:unconstrainedDPC} for the low-dimensional state-space example discussed in Section~\ref{sec:example}. The predictions blindly follow $\ybs_\text{ref}$ (independent of $\xib, \ubs$) for $\lambda = 0$ and tend towards $\hat\ybs_\text{LS}(\xib,\ubs)$ for $\lambda\to\infty$. Furthermore, $\hat\ybs_\text{DPC}(\xib, \ubs)$ is generally non-causal and time-variant (which cannot be observed in Figure~\ref{fig:unconstrainedDPC} due to the limited prediction horizon $N_f=1$). For finite $\lambda$ and $\ybs_\text{ref}\neq \zerob$, it is also an affine predictor (otherwise linear), regardless of the true system class generating the data. This last point highlights the interpretation of implicit predictors being ``the predictive behavior implicitly attributed to the data-generating system by the DPC scheme'' \cite{KLAEDTKE2023}, which may not necessarily match the true system if the scheme is poorly chosen.

\section{Effects of modifications in DPC}\label{sec:extensions}

While the previous section introduced the tools, namely trajectory-specific effect of regularization and implicit predictors, and exemplified them via results from \cite{KLAEDTKE2023, Klaedtke2024FCS_DPC, klaedtke2024unifyingPreprint}, we now extend these results towards some common modifications in DPC. We analyze the extension to affine DPC in Section~\ref{sec:affineDPC}, the inclusion of an offset in the regularization in Section~\ref{sec:offset}, the inclusion of slack variables in Section~\ref{sec:Slack}, and the inclusion of additional (terminal) equality constraints in Section~\ref{sec:terminalConstraints}. Finally, we also give two brief results on (recursive) feasibility for regularized DPC in Section~\ref{sec:feasibility}.

\subsection{DPC for affine systems}\label{sec:affineDPC}

While standard (linear) DPC can yield exact predictions for deterministic LTI systems (see the discussion in Section~\ref{sec:fundamentals}), exact extensions to particular classes of nonlinear systems have been proposed, e.g., in \cite{Berberich2020, Alsalti2021, Berberich2021_affineDPCexperimental}. Among these, we want to briefly discuss the case of affine time-invariant (ATI) systems 
\begin{subequations}\label{eq:statespace_ATI}
\begin{align}
    \xb (k+1) &= \Ab\xb(k)+\Bb\ub(k)+\eb\\
    \yb(k) &= \Cb\xb(k)+\Db\ub(k)+\rb.
\end{align}
\end{subequations}
proposed in \cite{Berberich2021_affineDPCexperimental}, which is also used for other nonlinear systems with continuously updated trajectory data in order to approximate a local (affine) linearization of the nonlinear system for predictions as in \cite{Berberich2021_affineDPCexperimental, Berberich2021}. 
Similarly to how trajectories of LTI systems \eqref {eq:statespace_LTI} can be generated by linear combinations, trajectories of ATI systems \eqref{eq:statespace_ATI} can be generated by affine combinations of trajectory data. That is, in addition to \eqref{eq:linearComb}, generated trajectories must also satisfy
\begin{equation}
    \sum_{i=1}^{\ell} \ab_i = \oneb_\ell^\top \ab = 1. \label{eq:affineCondition}
\end{equation}
Intuitively, this condition can be explained by noting that the effect of $\eb, \rb$ is present exactly once in each data trajectory and, accordingly, should be present exactly once in the generated trajectory.
Assuming exact data generated by an ATI system, the affine hull $\mathrm{aff}(\Dbc)$ of data columns in $\Dbc$ is equivalent to the set of all possible system trajectories iff \cite{Padoan2023}
\begin{equation}\label{eq:GPEaffin}
    \rank{\begin{matrix}
        \oneb_\ell^\top \\
        \Dbc
    \end{matrix}}=L m + n + 1. 
\end{equation} 
However, in the presence of noise and (other) nonlinearities, the same discussions as in Section~\ref{sec:fundamentals} apply, i.e., the unique and exact predictions are no longer possible. Accordingly, our analysis is not confined to affine DPC applied to data from ATI systems \eqref{eq:statespace_ATI}, but rather extends to affine DPC with data generated by any system, including the nonlinear tracking case in \cite{Berberich2021}. To understand the features of affine DPC in the presence of such realistic data, we extend our results from the linear DPC case.
As discussed in \cite[Sec.~II]{Mattsson2024} and, in particular, \cite[Rem.~4]{Mattsson2024}, many analysis results for linear DPC also apply to such nonlinear systems, which are linear in known (nonlinear) transformations of the state, input, and output. This also applies to the affine DPC scheme at hand, where we can simply consider $\check \xib := \begin{pmatrix}
    1 & \xib^\top
\end{pmatrix}^\top$ to be a nonlinear transformation of $\xib$, and $\widecheck \Wb := \begin{pmatrix}
    1_\ell & \Wb^\top
\end{pmatrix}^\top$ contains the corresponding data. Similarly, we define 
$$\check \zb := \begin{pmatrix}
    \check \xib \\ \ubs
\end{pmatrix} \quad \text{and} \quad \widecheck \Zb := \begin{pmatrix}
    \widecheck\Wb \\ \Ub
\end{pmatrix}$$ for the transformed state-input data. Instead of the linear least-squares estimates $\hat\ybs_\text{LS}(\xib, \ubs), \hat\ubs_\text{LS}(\xib)$, consider their affine counterparts
\begin{align*}
    \hat\ybs_\text{ALS}(\xib, \ubs) &:= \widecheck\Gb_\text{ALS} \check\zb = \Gb_\text{ALS} \zb + \gb_\text{ALS}\\
    \hat\ubs_\text{ALS}(\xib) &:= \widecheck\Kb_\text{ALS} \check\xib = \Kb_\text{ALS} \xib + \kb_\text{ALS}
\end{align*}
with 
    \begin{align*}
    \widecheck\Gb_\text{ALS} &:= \argmin_\Gb \|\Yb - \Gb \widecheck\Zb\|_\mathrm{F}^2 = \Yb \widecheck\Zb^+ =\begin{pmatrix}\gb_\text{ALS} & \Gb_\text{ALS}  \end{pmatrix} \\
    \widecheck\Kb_\text{ALS} &:= \argmin_\Kb \|\Ub - \Kb \widecheck\Wb\|_\mathrm{F}^2 = \Ub \widecheck\Wb^+ =\begin{pmatrix}\kb_\text{ALS} & \Kb_\text{ALS}  \end{pmatrix},
\end{align*}
and the corresponding residual matrices
$$
    \widecheck\dYb := \Yb - \widecheck\Gb_\text{ALS} \widecheck\Zb \quad \text{and} \quad 
    \widecheck\dUb := \Ub - \widecheck\Kb_\text{ALS} \widecheck\Wb.
$$
The analysis of affine DPC with regularizations in terms of their trajectory-specific effect and implicit predictors then follows accordingly. In the following, we present the case of (projection-based) quadratic regularization.
\begin{proposition}\label{prop:aff2NormCost}
    For affine DPC, the trajectory-specific effect of $h(\ab) = \lambda \|\ab\|_2^2$ is given by \begin{subequations}\label{eq:aff2NormCost}
        \begin{align}
        \lambda\|\ab^\ast(\xib, \ubs, \ybs)\|_2^2 &= \lambda \|\ybs-\hat\ybs_\text{ALS}(\xib, \ubs)\|_{\widecheck\Qbc_\text{reg}}^2 \label{eq:aff2NormCostTerm1}\\
    &\qquad + \lambda \left\|\ubs-\hat\ubs_\text{ALS}\left(\xib\right) \right\|_{\widecheck\Rbc_\text{reg}}^2 \label{eq:aff2NormCostTerm2}\\
    & \qquad+ \lambda \left\|\widecheck\xib\right\|_{\left(\widecheck\Wb\widecheck\Wb^\top\right)^{-1}}^2 \label{eq:aff2NormCostTerm3}
    \end{align}
    \end{subequations}
    with weighing matrices 
    $$
        \widecheck\Qbc_\text{reg} := \left(\widecheck\dYb\widecheck\dYb^\top\right)^{-1} \;\; \text{and} \;\;  \Rbc_\text{reg} := \left(\widecheck\dUb\widecheck\dUb^\top\right)^{-1}.
    $$
\end{proposition}
\begin{proof}
    The proof follows analogously to the linear DPC case.
\end{proof}
The interpretation of these cost terms also directly follows from the discussion below \eqref{eq:2NormCost}. Importantly, note that instead of the least-squares estimates for a linear predictor/controller $\hat\ybs_\text{LS}(\xib, \ubs), \hat\ubs_\text{LS}(\xib)$ the regularization now favors the least-squares estimates for the affine predictor/controller $\hat\ybs_\text{ALS}(\xib, \ubs), \hat\ubs_\text{ALS}(\xib)$ and shifts the predicted trajectory towards them. Similarly to \eqref{eq:proj2NormCost}, the effect of the cost terms \eqref{eq:aff2NormCostTerm2} and \eqref{eq:aff2NormCostTerm3} can be eliminated by considering a projection-based quadratic regularization $h(\ab) = \lambda\|\widecheck\Pib_\perp \ab\|_2^2$ with
$$
    \widecheck\Pib := \widecheck\Zb^+ \widecheck\Zb \quad \text{and} \quad \widecheck\Pib_\perp := \Ib - \widecheck\Pib.
$$
Finally, the analysis of predictive behavior via implicit predictors follows accordingly.
\begin{proposition}
\label{prop:implicitPredictorAffin}
    Consider affine DPC with quadratic output-tracking objective \eqref{eq:outputTrackingObj}, (projection-based) quadratic regularization $h(\ab)=\lambda\|\ab\|_2^2$ or $h(\ab)=\lambda\|\widecheck\Pib_\perp\ab\|_2^2$, and without additional output constraints. Under Assumption~\ref{assum:fullRank}, 
\begin{align}\label{eq:implicitPredictorADPC}
        \hat\ybs_\text{ADPC}(\xib, \ubs) &= \left(\Qbc + \lambda \widecheck\Qbc_\text{reg}\right)^{-1} \lambda \widecheck\Qbc_\text{reg} \hat \ybs_\text{ALS}(\xib, \ubs) \\
        & \quad + \left(\Qbc + \lambda \widecheck\Qbc_\text{reg}\right)^{-1} \Qbc \ybs_\text{ref} \nonumber
    \end{align}
    is an implicit predictor for this problem.
\end{proposition}
\begin{proof}
    The proof follows analogously to the linear DPC case from the trajectory-specific effect of regularization in \eqref{eq:aff2NormCost}.
\end{proof}
Again, the interpretation of this predictor follows from the linear DPC case discussed below \eqref{eq:DPC_predictor}.

\subsection{Regularization with offset}\label{sec:offset}

Some DPC schemes, in which the tracking of a non-zero equilibrium is desired, modify the regularization by including an offset, i.e., $\tilde h(\ab) = h(\ab-\overline{\ab})$ \cite{Berberich2021, Elokda2021QuadrotorDPC}. Although this adjustment is based on the assumption that the original regularization shifts predicted trajectories toward zero (see our discussion in Section~\ref{sec:motivation}), it introduces some interesting features, which can be analyzed via its trajectory-specific effect. Although \cite{Elokda2021QuadrotorDPC} considers various norm choices, we focus on the quadratic regularization $h(\ab) = \lambda\|\ab-\overline\ab\|_2^2$ used in both \cite{Berberich2021} and \cite{Elokda2021QuadrotorDPC}. Additionally, while the results in \cite{Berberich2021} are framed for an affine DPC setting, our analysis will focus on linear DPC as in \cite{Elokda2021QuadrotorDPC}. Note that these results can be readily extended to the affine case, as discussed in Section~\ref{sec:affineDPC}. Analogously to how $\ab$ generates the trajectory $\wb=\Dbc\ab$, we denote the trajectory generated by $\overline\ab$ as $\overline\wb:=\Dbc\overline\ab$, and similarly with $\overline\zb, \overline\ubs, \overline\ybs, \overline\xib$. Furthermore, we denote the difference between the two as 
$$
    \Delta\wb:=\Dbc\ab-\Dbc\overline\ab=\wb-\overline\wb,
$$
and similarly with $\Delta\zb, \Delta\ubs, \Delta\ybs, \Delta\xib$.
\begin{theorem}
    Consider the DPC problem \eqref{eq:DPC} with quadratic offset-regularization $h(\ab) = \lambda\|\ab-\overline\ab\|_2^2$. Under Assumption~\ref{assum:fullRank}, the trajectory-specific effect of this regularization is given by
    \begin{subequations}\label{eq:2NormOffsetCost}
    \begin{align}
        \lambda\|\ab^\ast(\xib, \ubs, \ybs)-\overline\ab\|_2^2 &= \lambda \|\Delta\ybs-\hat\ybs_\text{LS}(\Delta\xib, \Delta\ubs)\|_{\Qbc_\text{reg}}^2 \label{eq:2NormOffsetCostTerm1}\\
    &\qquad + \lambda \|\Delta\ubs-\hat\ubs_\text{LS} (\Delta\xib)\|_{\Rbc_\text{reg}}^2 \label{eq:2NormOffsetCostTerm2}\\
    & \qquad+ \lambda\|\Delta\xib\|_{\left(\Wb\Wb^\top\right)^{-1}}^2. \label{eq:2NormOffsetCostTerm3}
\end{align}
\end{subequations}
with $\Qbc_\text{reg}$ and $\Rbc_\text{reg}$ as defined in \eqref{eq:regWeights}.
\end{theorem}
\begin{proof}
    For a given tuple $(\xib, \ubs, \ybs, \overline\ab)$, one can check that the parametric optimal solution to \eqref{eq:OP_trajectorySpecific} is given by
    \begin{align*}
    \ab^\ast(\xib, \ubs, \ybs, \overline\ab) &= \Dbc^+ \wb + \left(\Ib - \Dbc^+\Dbc\right)\overline\ab \\
    \lambda\|\ab^\ast(\xib, \ubs, \ybs, \overline\ab)-\overline\ab\|_2^2 &= \lambda \Delta\wb^\top\left(\Dbc \Dbc^\top\right)^{-1}\Delta\wb.
    \end{align*}
    Except for the translation into $\Delta$-coordinates, this expression is equivalent to the one obtained in \cite[Sec.~III.A]{KLAEDTKE2023}, and the same block-matrix decomposition steps of $\left(\Dbc \Dbc^\top\right)^{-1}$ lead to \eqref{eq:2NormOffsetCost}.
\end{proof}
Hence, an offset by $\overline\ab$ in the regularization simply translates to the same trajectory-specific effect with an offset $\overline\wb$ for the considered trajectory. Note that \eqref{eq:2NormOffsetCost} can also be equivalently expressed as
\begin{align}
    \lambda \|\ab^\ast(\xib, \ubs, \ybs, \overline\ab) \! - \!\overline\ab\|_2^2 \! &= \! \lambda \|\ybs  \! - \!\hat\ybs_\text{LS}(\xib, \ubs) \! - \! \left(\overline\ybs \! - \!\hat\ybs_\text{LS}\left(\overline\xib, \overline\ubs\right)\right)\!\|_{\Qbc_\text{reg}}^2 \nonumber\\
    &\quad + \lambda \|\ubs  \! - \!  \hat\ubs_\text{LS}(\xib) \! - \!  \left(\overline\ubs-\hat\ubs_\text{LS}\left(\overline\xib\right)\right)\|_{\Rbc_\text{reg}}^2 \nonumber\\
    & \quad+ \lambda\|\xib  \! - \!  \overline \xib\|_{\left(\Wb_p\Wb_p^\top\right)^{-1}}^2. \label{eq:offsetAlternativeCost}
\end{align}
 While the third cost term is always irrelevant, we can see that the first two terms also may have no additional effect (compared to the usual quadratic regularization) if the trajectory $\overline\wb$ adheres to 
$$
    \overline \ybs = \hat\ybs_\text{LS}(\overline\xib, \overline\ubs), \quad \text{or} \quad 
    \overline\ubs = \hat\ubs_\text{LS}\left(\overline\xib\right),
$$
respectively. Similarly to \eqref{eq:proj2NormCost}, the effect of the cost terms \eqref{eq:2NormOffsetCostTerm2} and \eqref{eq:2NormOffsetCostTerm3} can be eliminated by considering a projection-based quadratic regularization. On that note, we briefly remark that both
\begin{align*}
    h(\ab) &= \lambda \|\Pib_\perp \left(\ab-\overline\ab\right)\|_2^2 = \lambda \left(\ab^\top \Pib_\perp \ab - 2 \overline\ab\Pib_\perp\ab + \overline\ab^\top \Pib_\perp\overline\ab \right)\; \text{or} \\
    h(\ab) &= \lambda \|\Pib_\perp \ab-\overline\ab\|_2^2 = \lambda \left(\ab^\top \Pib_\perp \ab - 2 \overline\ab\Pib_\perp\ab + \overline\ab^\top \overline\ab \right),
\end{align*}
yield the same effect, since they only differ in a constant term unrelated to $\ab$. Furthermore, using the first term of the alternative cost expression \eqref{eq:offsetAlternativeCost}, we can also state an implicit predictor as follows.
\begin{proposition}
\label{prop:implicitPredictorOffset}
    Consider the DPC scheme \eqref{eq:DPC} with quadratic output-tracking objective \eqref{eq:outputTrackingObj}, (projection-based) quadratic offset-regularization $h(\ab)=\lambda\|\ab-\overline\ab\|_2^2$ or $h(\ab)=\lambda\|\Pib_\perp\ab-\overline\ab\|_2^2$, and without additional output constraints. Under Assumption~\ref{assum:fullRank}, 
\begin{align}
        \hat\ybs_\text{DPC,off}(\xib, \ubs)=&
        \left(\lambda \Qbc_\text{reg}+\Qbc\right)^{-1} \lambda \, \Qbc_\text{reg}\, \hat\ybs_\text{LS,off}(\xib,\ubs) \nonumber\\
        &\;+\left(\lambda \Qbc_\text{reg}+\Qbc\right)^{-1} \Qbc \, \ybs_\text{ref}. \nonumber
\end{align}
with
\begin{equation}\label{eq:yLS_off}
    \hat\ybs_\text{LS,off}(\xib,\ubs) =   \hat\ybs_\text{LS}(\xib, \ubs) + \overline\ybs -\hat\ybs_\text{LS}\left(\overline\xib, \overline\ubs\right)
\end{equation}
    is an implicit predictor for this problem.
\end{proposition}
\begin{proof}
    The proof follows analogously to the linear DPC case from the trajectory-specific effect of regularization in \eqref{eq:offsetAlternativeCost}.
\end{proof}
We want to highlight that $\hat\ybs_\text{LS,off}(\xib,\ubs)$ in \eqref{eq:yLS_off} is an affine predictor mapping, which simply results from shifting the linear least-squares predictor $\hat\ybs_\text{LS}(\xib, \ubs)$ by a constant term $\overline\ybs -\hat\ybs_\text{LS}\left(\overline\xib, \overline\ubs\right)$, which depends on the discrepancy between said offset trajectory $\overline\wb$ and the least-squares predictor.
Other than that, the interpretation of this predictor follows from the linear DPC case discussed below \eqref{eq:DPC_predictor}. 

\subsection{Effect of slack variables in DPC}\label{sec:Slack}

A common modification to DPC, first introduced in \cite{Coulson2019DeePC}, is the inclusion of a slack variable as follows
\begin{subequations}
\label{eq:DPC_withSlack}
\begin{align}
\min_{\ubs,\ybs,\ab} 
 J(\xib, \ubs, &\ybs) + h(\ab) + h_\sigmab(\sigmab) \label{eq:DPCcostC_withSlack} \\
\text{s.t.} \quad \quad  \begin{pmatrix}
     \ubs_p \\ \ybs_p \\ \ubs \\ \ybs 
\end{pmatrix} + \begin{pmatrix}
    \zerob \\ \sigmab \\ \zerob \\ \zerob
\end{pmatrix} &= \begin{pmatrix}
    \Ub_p \\ \Yb_p \\ \Ub \\ \Yb 
\end{pmatrix}\ab, \label{eq:DPCeqConstrC_withSlack}\\
\left(\ubs, \ybs \right) &\in \Uc \times \Yc.  \label{eq:DPCsetConstrC_withSlack}
\end{align}
\end{subequations}
Here, we briefly decompose the notation from $\xib, \Wb$ back to $\ubs_p, \ybs_p, \Ub_p, \Yb_p$ to show exactly which part the slack variable $\sigmab$ is acting on. The use of slack variables is common to avoid infeasibility of the initial condition $\Wb\ab = \xib$ in situations where $\Wb$ does not have full row-rank. That is, while in the deterministic LTI case $\xib \in \image{\Wb}$ is guaranteed by the system dynamics, even a tiny amount of noise or nonlinearity may cause a measured initial condition $\xib$ to become inconsistent with $\image{\Wb}$. Note that some works like \cite{Berberich2021} extend the use of slack variables to all output-related parts of the image representation, i.e., they add another slack variable to the last block-row of \eqref{eq:DPCeqConstrC_withSlack}, and our analyses will naturally extend to those cases as well. In order to keep the use of $\sigmab$ small, another regularization term $h_\sigmab(\sigmab)$ is added to the objective function, where typical choices are $h_\sigmab(\sigmab) = \lambda_\sigmab\|\sigmab\|_1$ \cite{Coulson2019DeePC} or $h_\sigmab(\sigmab) = \lambda_\sigmab\|\sigmab\|_2^2$ \cite{Berberich2021}, and we will focus on the latter in combination with quadratic regularization $h(\ab) = \lambda \|\ab\|_2^2$.
While adding slack variables is not strictly necessary in cases where Assumption~\ref{assum:fullRank} applies, we will show an alternative interpretation of slack variables, which actually helps further justifying Assumption~\ref{assum:fullRank}, instead.
Consider the re-scaling
$$
\tilde\sigmab := \sqrt{\frac{\lambda_\sigma}{\lambda}} \sigmab 
$$
for which we have $\lambda_\sigma \|\sigma\|_2^2=\lambda \|\tilde\sigma\|_2^2$. Furthermore, we can rearrange \eqref{eq:DPCeqConstrC_withSlack} as 
$$
    \begin{pmatrix}
    \ubs_p  \\ \ybs_p \\ \ubs \\ \ybs 
\end{pmatrix}\! = \! 
\begin{pmatrix}
    \Ub_p & \zerob \\
    \Yb_p &  -\sqrt{\frac{\lambda}{\lambda_\sigma}} \Ib_{p \cdot N_p} \\    

    \Ub & \zerob  \\ 
    \Yb & \zerob 
\end{pmatrix}
\begin{pmatrix}
    \ab \\ \tilde\sigmab 
\end{pmatrix}\!
=: \begin{pmatrix}
    \widetilde{\Wb} \\ \Ub \\ \Yb
\end{pmatrix} \tilde\ab
=: \widetilde{\Dbc} \tilde\ab
$$
and sum up the regularization terms 
$$
\lambda_\sigma \|\sigma\|_2^2 + \lambda \|\ab\|_2^2 = \lambda \|\tilde \ab\|_2^2:= h(\tilde\ab).
$$
Using this re-parameterization, the slack variables can be interpreted as adding artificial trajectory data columns in the new augmented data matrix $\widetilde\Dbc$, where $\widetilde{\Wb}$ is now guaranteed to be full row-rank under the usual input persistency of excitation conditions (e.g., the one used in \cite[Thm.~1]{Berberich2020stability}), even if $\Wb$ is not. If an additional slack variable is used for the last block-row of future outputs $\ybs$ (like in \cite{Berberich2020stability, alsalti2024robustDPC}), the same input persistency of excitation conditions guarantee $\widetilde\Dbc$ to be full row-rank, i.e., Assumption~\ref{assum:fullRank} to be satisfied, even if $\Dbc$ is not. Hence, the consideration of slack variables further increases the scope of cases, where the (already commonly made) Assumption~\ref{assum:fullRank} is justified. Other than that, all analyses made using the trajectory-specific effect of regularization and implicit predictors for DPC with quadratic regularization $h(\ab) = \lambda \|\ab\|_2^2$ carry over by simply considering $\widetilde\Dbc$ instead of $\Dbc$. The same does not apply to projection-based regularization $h(\ab) = \lambda \|\Pib_\perp \ab\|_2^2$, since the combined weight matrix $\Sb = \diag{\Pib_\perp, \Ib}$ for the resulting $h(\tilde\ab) = \lambda \|\Sb \tilde\ab\|_2^2$ has not yet been analyzed in the literature. For future work, we consider a more general analysis of quadratic regularization $h(\ab) = \lambda \|\Sb\ab\|_2^2$ with arbitrary weight matrix $\Sb$ an insightful endeavor.

Regarding the tuning of $\lambda_\sigmab$ with respect to $\lambda$, how prominently the artificial trajectories are used in the resulting DPC predictions mainly depends on the ratio $\lambda \lambda_\sigma^{-1}$ and its ratio w.r.t. the signal power of the data contained in $\Yb_p$. Empirically, one can also observe an improved conditioning of the matrix $\widetilde\Wb\widetilde\Wb^\top$ compared to $\Wb\Wb^\top$ for increasing $\lambda \lambda_\sigma^{-1}$, the quantification of which might also be interesting for future work. An interesting (and perhaps unexpected) link shows up in \cite{Yin2023}, where the authors iteratively optimize the weight $\lambda$ for $\lambda_\sigma=1$ (see \cite[Eq.~(33)]{Yin2023}) in order to compute an approximation to the maximum-likelihood predictor for data generated by an LTI system with additive output noise.

\subsection{Predictive behavior with (terminal) equality constraints}\label{sec:terminalConstraints}

In order to provide closed-loop stability guarantees, classical MPC typically makes use of terminal ingredients (see, e.g., \cite{Mayne2000}). Similarly, practical stability for some DPC schemes like \cite{Berberich2020stability, Berberich2021, alsalti2024robustDPC} has been proven by employing (among other modifications) terminal equality constraints for the last steps of the predicted I/O sequence (see \cite{Berberich2022Tutorial} for a tutorial). While we view terminal constraints as the main use case of our following analysis, the results naturally expand to other kinds of equality constraints. When analyzing the effect of additional terminal equality constraints in the DPC scheme \eqref{eq:DPC}, first, let us briefly recall that the trajectory-specific effect of regularization introduced in Section~\ref{sec:trajectory-specific} is universally unaffected, since it applies to any chosen triple $(\xib, \ubs, \ybs)$ satisfying \eqref{eq:OP_trajectorySpecific_constr}, and thus also the ones satisfying additional (terminal equality) constraints. Regarding the characterization of predictive behavior via implicit predictors, similar considerations apply for the effect of additional constraints on the input sequence $\ubs$, as already discussed below \eqref{eq:implicitPredictorOP}.

However, the predictive behavior of DPC is significantly influenced by any kind of output constraints and often contradicts the unconstrained behavior as observed in \cite[Sec.~III.C]{KLAEDTKE2023}. To simplify upcoming notation, we assume that the terminal output constraints require the predicted output $\ybs$ to match the reference $\ybs_\text{ref}$ over the final $n$ (or any other amount of) steps of the prediction. This assumption is made without loss of generality; if the original reference $\ybs_\text{ref}$ does not naturally satisfy this condition, we can simply define a modified reference $\tilde{\ybs}_\text{ref}$, where the last $n$ steps are adjusted to incorporate the terminal constraints. Once again, we make use of Assumption~\ref{assum:fullRank}, which is further justified by the use of slack variables in \cite{Berberich2020stability, alsalti2024robustDPC}, as discussed in Section~\ref{sec:Slack}. Furthermore, we assume 
\begin{equation}\label{eq:defQ}
    \Qbc := \diag{\Qb, \hdots, \Qb}\; \text{with positive definite}\; \Qb\in \R^{p\times p},
\end{equation}
which aligns with the setup in \cite{Berberich2020stability}. 
Given these assumptions, the following theorem precisely characterizes the effect of (terminal) output equality constraints on the predictive behavior of DPC with (projection-based) quadratic regularization.

\begin{figure}
    \centering
        \includegraphics[trim=5.6cm 12.4cm 5.92cm 12.4cm,clip=true, scale=0.87]{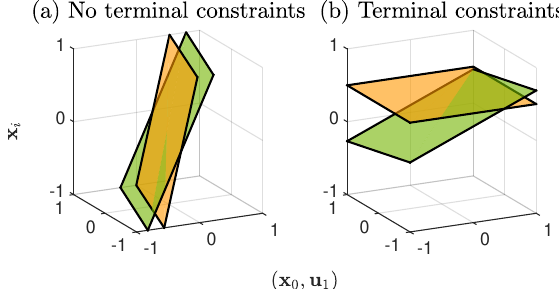}
        \caption{Implicit predictor $\hat\xbs_\text{DPC}(\xb_0, \ubs)$ for the DPC example discussed in Section~\ref{sec:example} with prediction horizon $N_f = 2$ and reference $\xbs_\text{ref} = \begin{pmatrix}
            0.5 & 0.5
        \end{pmatrix}^\top$. To deal with the higher dimensionality, the first (green) and second (orange) prediction step of $\hat\xbs_\text{DPC}(\xb_0, \ubs)$ are visualized individually. Furthermore, we constrain the second input to $\ubs_2=0$ and only visualize the first input $\ubs_1$.}
        \label{fig:terminalConstrainedDPC}
\end{figure}

\begin{theorem}\label{thm:implicitPredictorEqConstr}
    Consider the DPC problem \eqref{eq:DPC} with $h(\ab) = \lambda \|\ab\|_2^2$ or $h(\ab) = \lambda \|\Pib_\perp\ab\|_2^2$, and the additional terminal equality constraint
    \begin{equation}\label{eq:terminalConstr}
        \begin{pmatrix}
            \zerob & \Ib_n
        \end{pmatrix}
        \ybs 
        = 
        \begin{pmatrix}
            \zerob & \Ib_n
        \end{pmatrix}
        \ybs_\text{ref}. 
    \end{equation}
    Furthermore, consider the block partitioning
    $$
        \Qbc = \begin{pmatrix}
            \Qbc_1 & \zerob \\
            \zerob & \Qbc_2
        \end{pmatrix}, \quad 
        \Qbc_\text{reg} = \begin{pmatrix}
            \Qbc_\text{reg}^{(11)} & \Qbc_\text{reg}^{(12)} \\
            \Qbc_\text{reg}^{(21)} & \Qbc_\text{reg}^{(22)}
        \end{pmatrix}
    $$
    with $\Qbc_2, \Qbc_\text{reg}^{(22)} \in \R^{np \times np}$ aligning with the assumption made on the structure of $\Qbc$ in \eqref{eq:defQ}. Under Assumption~\ref{assum:fullRank}, 
    \begin{equation}\label{eq:implicitPredictorADPC_eqConstr}
        \hat\ybs_\text{DPC}(\xib, \ubs) = \Lambda_\text{ref} \; \ybs_\text{ref} +  \Lambda_\text{reg} \; \hat \ybs_\text{LS}(\xib, \ubs) 
    \end{equation}
    with the weights
    { \begin{align*}
    \Lambda_\text{ref} \!&:=\!
        \begin{pmatrix}
             \!\left( \!\Qbc_1 \! + \! \lambda  \Qbc_\text{reg}^{(11)} \!\right)^{-1} \!\! \Qbc_1 
             &  -\left(\!\Qbc_1 \! + \! \lambda  \Qbc_\text{reg}^{(11)}\!\right)^{-1}\!\! \lambda \Qbc_\text{reg}^{(12)}\! \\
            \zerob & \Ib
        \end{pmatrix} \\
    \Lambda_\text{reg} \!&:=\!
        \begin{pmatrix}
            \!\left(\!\Qbc_1 \! + \! \lambda  \Qbc_\text{reg}^{(11)}\!\right)^{-1}\!\! \lambda  \Qbc_\text{reg}^{(11)} 
            & \!\!\left(\!\Qbc_1 \! + \! \lambda  \Qbc_\text{reg}^{(11)}\!\right)^{-1}\!\! \lambda \Qbc_\text{reg}^{(12)}\! \\
            \zerob & \zerob
        \end{pmatrix}   
    \end{align*}}
    is an implicit predictor for this problem.
\end{theorem}
\begin{proof}
    The proof strategy lies in characterizing the effect of hard terminal constraints as soft constraints with costs tending to infinity. Note that this is an exact characterization that works due to Assumption~\ref{assum:fullRank} allowing for any triple $(\xib, \ubs, \ybs)$ in \eqref{eq:DPCeqConstr} and would be invalid for an indirect (i.e., model-based) scheme, where the set of feasible $(\xib, \ubs, \ybs)$ is limited by a prediction model enforced as a hard constraint.
    The result thus follows by considering $\Qbc_q := \diag{\Qbc_1, q \Ib}$ instead of $\Qbc$ in \eqref{eq:DPC_predictor} and taking the limit $q \to \infty$. For brevity, we only show the limit computation 
    $$
        \Lambda_\text{ref} = \lim_{q\to \infty} \left(\lambda \Qbc_\text{reg} + \Qbc_q\right)^{-1}\Qbc_q,
    $$
    and note that 
    $$
        \Lambda_\text{reg} = \lim_{q\to \infty} \left(\lambda \Qbc_\text{reg} + \Qbc_q\right)^{-1} \lambda \Qbc_\text{reg}
    $$
    follows analogously. From the structure and positive definiteness of $\Qbc_q$ we have $\Qbc_q^{-1} = \diag{\Qbc_1^{-1}, q^{-1} \Ib}$, and therefore
    \begin{align*}
        \left(\lambda \Qbc_\text{reg} + \Qbc_q\right)^{-1}\Qbc_q &= \left(\lambda\Qbc_q^{-1} \Qbc_\text{reg} + \Ib\right)^{-1} \\
       &= \begin{pmatrix}
            \Ib+\Qbc_1^{-1} \lambda \Qbc_\text{reg}^{(11)} & \!\Qbc_1^{-1} \lambda \Qbc_\text{reg}^{(12)} \\
            q^{-1} \lambda \Qbc_\text{reg}^{(21)} & \! \Ib + q^{-1} \lambda \Qbc_\text{reg}^{(22)} 
        \end{pmatrix}^{-1}\!\!\!.
    \end{align*}
    Computing the limit for this expression yields
    \begin{align*}
        &\lim_{q\to\infty} \left(\lambda \Qbc_\text{reg} + \Qbc_q\right)^{-1}\Qbc_q \\
        =& 
        \begin{pmatrix}
            \Ib +\Qbc_1^{-1} \lambda \Qbc_\text{reg}^{(11)} & \Qbc_1^{-1} \lambda \Qbc_\text{reg}^{(12)} \\
            \zerob & \Ib
        \end{pmatrix}^{-1}\\
        =& 
        \begin{pmatrix}
            \left(\Ib+\Qbc_1^{-1} \lambda \Qbc_\text{reg}^{(11)}\right)^{-1} & \!\! -\left(\Ib+\Qbc_1^{-1} \lambda \Qbc_\text{reg}^{(11)}\right)^{-1} \!\! \Qbc_1^{-1} \lambda \Qbc_\text{reg}^{(12)} \\
            \zerob & \Ib
        \end{pmatrix} \\
        =& 
        \begin{pmatrix}
            \left(\Qbc_1+ \lambda \Qbc_\text{reg}^{(11)}\right)^{-1} \Qbc_1 &  -\left(\Qbc_1+ \lambda \Qbc_\text{reg}^{(11)}\right)^{-1}  \lambda \Qbc_\text{reg}^{(12)} \\
            \zerob & \Ib
        \end{pmatrix},
    \end{align*}
    where we used a well known block-matrix inversion formula in the third step. The computation of $\Lambda_\text{reg}$ follows analogously.
\end{proof}
As expected, the last $n$ steps of the predicted $\hat\ybs_\text{DPC}(\xib, \ubs)$ fully align with the terminal equality constraints given by \eqref{eq:terminalConstr}, which can be seen from the zero and identity blocks in the lower block-row of $\Lambda_\text{ref}$ and $\Lambda_\text{reg}$.
 Furthermore, it is remarkable that the structure of a (matrix-)weighed sum observed in \eqref{eq:DPC_predictor} is preserved in the upper blocks of $\Lambda_\text{ref}$ and $\Lambda_\text{reg}$, characterizing how the trade-off between $\ybs_\text{ref}$ and $\hat \ybs_\text{LS}(\xib, \ubs)$ in the first $N_f-n$ prediction steps is affected by enforcing equality to the reference for the last $n$ steps. This trade-off is visualized in Figure~\ref{fig:terminalConstrainedDPC} for the low-dimensional state-space example discussed in Section~\ref{sec:example}. This visualization also highlights the discrepancy between the effect of terminal constraints in regularized DPC compared to model-based schemes, where a predictor is enforced as hard constraints $\ybs=\hat\ybs(\xib, \ubs)$. In the latter, a contradiction between $\hat\ybs(\xib, \ubs)$ and the constraint \eqref{eq:terminalConstr} would lead to infeasibility of the corresponding $(\xib, \ubs)$. Instead, regularized DPC not only changes the last $n$ steps of predictions to match \eqref{eq:terminalConstr}, but alters the first $N_f-n$ steps as specified in Theorem~\ref{thm:implicitPredictorEqConstr}. While this preserves the feasibility, which we next discuss in Section~\ref{sec:feasibility}, it may be an unintuitive and potentially unwanted feature for practitioners who are unaware of it.

\subsection{(Recursive) feasibility in DPC}\label{sec:feasibility}

Recursive feasibility of the closed-loop is an important concept in stabilizing predictive control schemes, and a lot of focus is placed on either guaranteeing it a priori or certifying it for a given controller \cite{LOFBERG2012}. Correspondingly, practical stability results of DPC schemes also include results on recursive feasibility (see \cite[Sec.~IV.D]{Berberich2020stability}, \cite[Prop.~IV.1]{Berberich2022Tutorial}, \cite[Thm.~14]{alsalti2024robustDPC}). Our work emphasizes the analysis of DPC schemes through the structure of the underlying OCP, without making assumptions about the class of systems generating the data. While we should not expect in-depth closed-loop analysis results without making such assumptions, our approach still reveals broad results on (recursive) feasibility based on the OCP structure, which seem to be currently overlooked in the literature. Although the DPC scheme \eqref{eq:DPC} does not include constraints $\xib \in \Xc$ on the initial state, we will consider them here to state more nuanced and general results.
\begin{proposition}\label{prop:feasibleSet}
    Consider the DPC scheme \eqref{eq:DPC} with non-empty constraint sets $\Uc, \Yc$, and additional state constraints $\xib \in \Xc$. Under Assumption~\ref{assum:fullRank}, the set of feasible initial states $\xib$ is equivalent to $\Xc$.
\end{proposition}
\begin{proof}
    Under Assumption~\ref{assum:fullRank}, there exists an $\ab$ satisfying \eqref{eq:DPCeqConstr} for any triple $(\xib, \ubs, \ybs)$. Hence, for any $\xib \in \Xc$ there also exist $(\ubs, \ybs)$ satisfying the remaining constraints \eqref{eq:DPCsetConstr}.
\end{proof}
The following is a simple consequence of this result.
\begin{proposition}\label{eq:recFeasibility}
    Consider the DPC scheme \eqref{eq:DPC} with non-empty constraint sets $\Uc, \Yc$, and without additional state constraints, i.e., $\xib \in \R^{(m+p)N_p}$. Under Assumption~\ref{assum:fullRank}, the OCP \eqref{eq:DPC} is always feasible (and hence recursively feasible).
\end{proposition}
\begin{proof}
    Simply consider $\Xc=\R^{(m+p)N_p}$ in the proof of Proposition~\ref{prop:feasibleSet}.
\end{proof}
Although straightforward and already briefly discussed in \cite[Sec.~III.C]{KLAEDTKE2023}, these results have important implications. The DPC problem \eqref{eq:DPC} is typically (i.e., under Assumption~\ref{assum:fullRank}) always feasible. This feasibility is by design of the OCP itself, rather than coming from the closed-loop control or system dynamics. For the schemes in \cite{Berberich2020stability, Berberich2022Tutorial}, the use of slack variables on the output variables $\ybs_p$ and $\ybs$ ensures that Assumption~\ref{assum:fullRank} holds for the extended data matrix $\widetilde\Dbc$ (see our discussion in Section~\ref{sec:Slack}) under typical input persistency of excitation assumptions \cite[Thm.~1]{Berberich2020stability}. Similarly, in \cite{alsalti2024robustDPC}, where slack variables are applied to the entire trajectory variable $\wb$, Assumption~\ref{assum:fullRank} is also always satisfied for $\widetilde\Dbc$. Consequently, the recursive feasibility analysis in these schemes could be replaced by Proposition~\ref{eq:recFeasibility}. Finally, we want to highlight the generality of these results. For example, the considered constraint sets $\Uc, \Yc$ may also include terminal equality constraints like the ones discussed in Section~\ref{sec:terminalConstraints}.

\section{Conclusions and Outlook}
\label{sec:ConclusionOutlook}

This work discussed the use of trajectory-specific effects of regularizations (see Definition~\ref{def:trajectorySpecific}) and implicit predictors (see Definition~\ref{def:implicit_predictor}) as analysis tools to improve explainability in regularized DPC. The former concretizes the effects of any regularization $h(\ab)$ by eliminating auxiliary variables and reformulating an equivalent cost $h^\ast(\xib, \ubs, \ybs)$, which is specific to the trajectory variables $(\xib, \ubs, \ybs)$, instead. The latter is a predictor mapping $\hat\ybs(\xib, \ubs)$, which coincides with the predictions generated while applying regularized DPC and thus provides an indirect (i.e., model-based) view on this direct scheme, as visualized in Figure~\ref{fig:implicitPredictor}. We first demonstrated the use of these tools by summarizing results from \cite{KLAEDTKE2023, Klaedtke2024FCS_DPC, klaedtke2024unifyingPreprint} on trajectory-specific effects and implicit predictors for (projection-based) quadratic regularization, and provided intuitive interpretations of the uncovered effects in Section~\ref{sec:trajectory-specific} and Section~\ref{sec:implicitPredictors}. We then extended these results towards common modifications in DPC, including DPC for affine systems, offset regularizations, slack variables, and terminal constraints. Notably, we showed that slack variables can be interpreted as introducing artificial data samples in an augmented data matrix $\widetilde\Dbc$, which can be analyzed just like before and is even more likely to satisfy the already common Assumption~\ref{assum:fullRank}.
For each modification, we computed the trajectory-specific cost or presented an implicit predictor if the modification led to a change in the regularization effect or predictive behavior.
 A key feature of our results is modularity, since they can be freely combined. For example, the nonlinear tracking scheme \cite{Berberich2021} uses an affine formulation, offset regularization, slack variables, and terminal equality constraints. Hence, a combination of results from Sections~\ref{sec:affineDPC}-\ref{sec:terminalConstraints} can be used to explain the interactions between control objective, constraints, and regularization in this scheme.
Finally, we gave two brief results on (recursive) feasibility, stating that common DPC formulations are often feasible by design, hence eliminating the need for a deeper analysis of recursive feasibility in those cases.

Although this work primarily focused on DPC with (projection-based) quadratic regularizations, we emphasize the broadness of our proposed analysis tools, which are (in principle) applicable to any choice of regularization $h(\ab)$. Therefore, similar analyses for more general quadratic regularization $h(\ab)=\lambda\|\Sb\ab\|_2^2$, 1-norm regularization $h(\ab) = \lambda \|\ab\|_1$ (which is sometimes used for nonlinear systems \cite{Coulson2019DeePC}), or general p-norms $h(\ab) = \lambda \|\ab\|_p$ should yield similarly insightful characterizations. Furthermore, similarly to the relation with $\gammab$-DDPC presented in Section~\ref{sec:isolatingEffects}, the trajectory-specific effect of regularizations may uncover links to other frameworks, such as generalized DPC \cite{Lazar2023GeneralizedDP}.
Finally, combining the generality of the presented analysis tools with more specific assumptions and prior knowledge about the true underlying system could help establish new robustness and stability results.

% References

\end{document}